\documentclass[reprint,aps,pra,rmp,superscriptaddress]{revtex4-1}
\usepackage{graphicx}
\usepackage{dcolumn}
\usepackage{bm}
\usepackage[latin5]{inputenc}
\usepackage{amssymb,amsmath,amsthm,amsbsy,mathptmx}
\DeclareMathAlphabet{\mathcal}{OMS}{cmsy}{m}{n}
\usepackage{tabularx}
\usepackage{xcolor}
\usepackage{mathtools}
\usepackage{oubraces}
\usepackage{dsfont}
\usepackage{graphicx}
\usepackage{rotating}
\usepackage[most]{tcolorbox}
\usepackage{enumerate}
\usepackage{subfloat}
\usepackage{cleveref}
\Crefname{subfigures}{figure}{figures}%
\Crefname{subfigures}{Figure}{Figures}%

\tcbset{colback=green!5!white, colframe=red!75!black,
        highlight math style= {enhanced, 
            colframe=red,colback=red!10!white,boxsep=0pt}
        }

\usepackage{hyperref}
\def\bra#1{\mathinner{\langle{#1}|}}
\def\ket#1{\mathinner{|{#1}\rangle}}
\def\braket#1{\mathinner{\langle{#1}\rangle}}

\begin{document}

\title{Deterministic transformations of coherent states under incoherent operations}
\author{G\"{o}khan Torun}
\email{torung@itu.edu.tr}
\affiliation{Department of Physics, Istanbul Technical University, Maslak 34469, Istanbul, Turkey}
\affiliation{Centre for the Mathematics and Theoretical Physics of Quantum Non-Equilibrium Systems,
School of Mathematical Sciences, The University of Nottingham, Nottingham NG7 2RD, United Kingdom}
\author{Ali Yildiz}
\affiliation{Department of Physics, Istanbul Technical University, Maslak 34469, Istanbul, Turkey}
\date{\today}


\begin{abstract}
It is well known that the majorization condition is the necessary and sufficient condition for the deterministic transformations of both pure bipartite entangled states by local operations and coherent states under incoherent operations. In this paper, we present two explicit protocols for these transformations. We first present a permutation-based protocol which provides a method for the single-step transformation of $d$-dimensional coherent states. We also obtain generalized solutions of this protocol for some special cases of $d$-level systems. Then, we present an alternative protocol where we use $d'$-level ($d'$ $<$ $d$) subspace solutions of the permutation-based protocol to achieve the complete transformation as a sequence of coherent-state transformations. We show that these two protocols also provide solutions for deterministic transformations of pure bipartite entangled states.
\end{abstract}

\maketitle


\section{Introduction}

Coherence, a core manifestation of nonclassicality, is used as a crucial resource in many quantum information processing tasks, such as quantum thermodynamics \cite{qthermody1,qthermody2,qthermody3,qthermody4}, quantum metrology \cite{qmetrology1,qmetrology2,qmetrology3,qmetrology4}, and quantum algorithms \cite{qalgorithms,qalgorithms1,qalgorithms2}.
As with all resources, coherence also needs to be quantified. In line with this goal, the relative entropy of coherence and the $l_1$ norm of coherence were presented in \cite{qcoherence} for the quantification of coherence as a resource. Manipulation of coherence is also an important part of resource theory of coherence \cite{resourcethercoh,Winter,Streltsovrtc,Dana,Yang,genuine,Alexrtofc}.  Quantum coherence has similar features with quantum entanglement \cite{entanglement}, the well-known fundamental resource in many quantum information processes, in the context of state-to-state transitions. Nielsen  \cite{condition} used the linear-algebraic theory of majorization and obtained the necessary and sufficient conditions for a class of entanglement transformations. It was later realized that it is possible, with the aid of a completely similar approach, to achieve the deterministic transformations of coherent states under incoherent operations. The researchers \cite{ccoherence,ccoherenceerr} have built the counterpart of Nielsen theorem for coherence manipulation and also showed that majorization is also a key ingredient for the interconvertibility of coherent states. An optimal local conversion strategy of bipartite entangled pure states was proposed by Vidal \cite{vidal}, which is also a generalization of
Nielsen's theorem \cite{condition}. This strategy was adapted to the optimal conversion of coherent states under incoherent operations \cite{ccoherence2}. See \cite{resourcethercoh} for a comprehensive review of the development of this rapidly growing research field that encompasses the characterization, quantification, manipulation, dynamical evolution, and operational application of quantum coherence.

While the interconversion of pure states under incoherent operations was studied in various papers \cite{ccoherence,ccoherence2,catalyticcoherence,Winter,Chitambar1,Chitambar,coherence-vector}, providing alternative and easily implementable protocols for the state-to-state transitions is of paramount importance in quantum resource theories. We will present two explicit protocols (followed by illustrative examples) for the deterministic transformations of coherent states via incoherent operations. One is a permutation-based protocol which provides the single-step transformation of $d$-level coherent states, and the other is a step-by-step transformation protocol. We use $d'$-level ($d' < d$) subspace solutions of the former to construct the latter, and for $d'=5$ the number of steps is $\lfloor {(d+2)}/{4} \rfloor$.

The paper is structured as follows. We begin with the summary of incoherent states, coherent states, and incoherent operations in Sec. II. This section ends with the definition of majorization criteria and an explanation of its connection with coherent states manipulation. In Sec. III, we construct two explicit protocols for the deterministic transformations of coherent states under incoherent operations. We explicitly examine illustrative examples with discussion for both protocols and present solutions for some generalizable cases
of the first protocol. At the end of Sec. III, we show how the protocols and solutions presented for coherence transformations also provide solutions for bipartite entangled pure state transformations. We conclude our work in Sec. IV.


\section{Preliminaries}

First, we define the basis-dependent notions of incoherent and coherent states. Quantum states that are diagonal with respect to a fixed orthonormal basis $\{\ket{i}\}_{i=1,2,\dots,d}$ are defined as incoherent, and they constitute a set labeled by $\mathcal{I}$ \cite{qcoherence}. All incoherent states $\rho \in \mathcal{I}$ are of the form
\begin{equation}\label{incoherent-state}
\rho=\sum_{i=1}^{d} p_i \ket{i}\bra{i},
\end{equation}
where $p_i\in [0,1]$ and $\sum_{i}p_i=1$. A finite $d$-dimensional pure coherent state, on the other hand, is given by
\begin{eqnarray}\label{psigenform}
\ket{\psi}=\sum_{i=1}^{d} \psi_i \ket{i}, \quad \psi_i \geq \psi_{i+1}> 0,
\end{eqnarray}
where $\{\psi_i\}_{i=1,2,\dots,d}$ are non-negative real numbers, and all complex phases have been eliminated by incoherent operations (diagonal unitaries),  satisfying $\sum_i \psi_i^2=1$.
The state given in Eq. \eqref{psigenform} is a maximally coherent state for $\{\psi_i\}_{i=1,\dots,d}=\{\frac{1}{\sqrt{d}}, \frac{1}{\sqrt{d}}, \dots, \frac{1}{\sqrt{d}}\}$, where all other $d$-dimensional coherent states can be generated from it by means of incoherent operations \cite{qcoherence}.

We focus on particular quantum operations for which measurement outcomes are retained as stated in \cite{qcoherence}. These quantum operations are defined by Kraus operators $\{K_i\}$ such that $\sum_{i}K_i^{\dag}K_i=I$ and, for all $i$ and $\rho \in \mathcal{I}$,
\begin{eqnarray}\label{freeoperation}
\rho \rightarrow \rho_i=\frac{K_i\rho K_i^{\dag}}{\text{Tr}[K_i\rho K_i^{\dag}]} \in \mathcal{I}.
\end{eqnarray}
Therefore, Kraus operators $\{K_i\}$ are called incoherent operators with the requirement $K_i\rho K_i^{\dag} \subset \mathcal{I}$ for all $i$, i.e., incoherent states remain incoherent under incoherent operations. The operations satisfying Eq. \eqref{freeoperation} are the free operations in the context of resource theories of coherence.

We now recall the majorization condition which plays a central role in the manipulation of bipartite pure entangled states and coherent states \cite{condition,NielsenVidal,ccoherence}.
Majorization is unambiguously defined in Chap. 2 of \cite{Bhatia}. Suppose $x\equiv (x_1,\dots, x_d)^T$ and  $y\equiv (y_1,\dots, y_d)^T$ are real $d$-dimensional vectors whose components are in decreasing order. Then $x$ is majorized by $y$ (equivalently $y$ dominates $x$), written $x \prec y$, if the inequalities
$\sum_{i=1}^{k}x_i \leq \sum_{i=1}^{k} y_i$
are satisfied for any $k\in [1,d-1]$ with equality holding when $k=d$.
Nielsen \cite{condition} showed that a bipartite entangled pure state $\ket{\Psi}=\sum_{i}\psi_i\ket{i}\ket{i}$ can be deterministically transformed by local operations and classical communication (LOCC) to another bipartite entangled pure state $\ket{\Phi}=\sum_{i}\phi_i\ket{i}\ket{i}$, whose Schmidt
coefficients are ordered in decreasing order, if and only if the vector $(\psi_1^2, \psi_2^2, \dots, \psi_d^2)^T$ is majorized by the vector $(\phi_1^2, \phi_2^2, \dots, \phi_d^2)^T$ written $(\psi_1^2, \dots, \psi_d^2)^T$ $\prec$ $(\phi_1^2, \dots, \phi_d^2)^T$.
Therefore, the majorization condition assures the transformation $\ket{\Psi} \rightarrow \ket{\Phi}$ deterministically for bipartite entangled pure states as part of entanglement manipulation. Majorization is also a good criterion that tells us whether one
state can be transformed into another under some incoherent operations. It was shown that a pure coherent state $\ket{\psi}$ given in Eq. \eqref{psigenform} can be deterministically transformed via incoherent operations  to another pure coherent state
\begin{eqnarray}\label{phigenform}
\ket{\phi}=\sum_{j=1}^{d} \phi_j \ket{j}, \quad \phi_j \geq \phi_{j+1}> 0,
\end{eqnarray}
if and only if the coherence vector, defined in \cite{coherence-vector},  $\mu{(\psi)}=(\psi_1^2, \psi_2^2, \dots, \psi_d^2)^T$ is majorized by the coherence vector $\mu{(\phi)}=(\phi_1^2, \phi_2^2, \dots, \phi_d^2)^T$ \cite{ccoherence}, written  $\mu{(\psi)}$ $\prec$  $\mu{(\phi)}$.  If the following inequalities,
\begin{eqnarray}\begin{aligned}\label{majorization}
\sum_{i=1}^{k}\psi_i^{2} &\leq \sum_{j=1}^{k}\phi_j^{2}
\end{aligned}\end{eqnarray}
are satisfied for any $k\in [1,d-1]$ with equality holding when $k=d$, i.e., $\sum_{i=1}^{d}\psi_i^2 = \sum_{j=1}^{d}\phi_j^2=1$, then  $\mu{(\psi)}$ $\prec$  $\mu{(\phi)}$.


\section{Deterministic Transformations of Coherent States Under Incoherent Operations}

In the following, we present two protocols for the deterministic transformations of the coherent state \eqref{psigenform} to the coherent state \eqref{phigenform} for which the majorization condition is satisfied. We reinforce them by discussing various examples. These two protocols also provide complete solutions for LOCC deterministic transformations between $d \otimes d$ bipartite entangled pure states.


\subsection{Protocol I}

\begin{table*}[]
\caption{All possible permutations (or unitary transformations) for $d$-dimensional coherent-state transformations via incoherent operations. The table contains permutations $\ket{x} \leftrightarrow \ket{y}$ where $y>x$. For a given case of $d$-level systems, the relation between the coefficients of the coherent states $\ket{\psi}=\sum_{j=1}^{d}{\psi_j}\ket{j}$ and $\ket{\phi}=\sum_{j=1}^{d}{\phi_j}\ket{j}$ is either $\psi_{k} \geq \phi_{k}$ or $\psi_{k} \leq \phi_{k}$ for any $k=2,3,...,d-1$ where $\psi_{1} \leq \phi_{1}$ and $\psi_{d} \geq \phi_{d}$ is satisfied due to the majorization condition. On the one hand, if the relation is $\psi_{k} \leq \phi_{k}$ then the row in which $\psi_{k} \geq \phi_{k}$ is located is crossed out. On the other hand, if the relation is $\psi_{k} \geq \phi_{k}$ then the column in which $\psi_{k} \leq \phi_{k}$ is located is crossed out.  In a table there will be a certain number, $\zeta$, of permutations where $(d-1)\leq\zeta\leq \lfloor d/2\rfloor \lceil d/2 \rceil$.}
\centering
\begin{ruledtabular}
\begin{tabular}{c c c c c c c c c}
& $\psi_1 \leq \phi_1$ & $\psi_2 \leq \phi_2$ & $\psi_3 \leq \phi_3$ & $\psi_4 \leq \phi_4$ & $\dots$ & $\psi_{d-2} \leq \phi_{d-2}$ & $\psi_{d-1} \leq \phi_{d-1}$ \\ [0.5ex] \hline
$\psi_2 \geq \phi_2$ & $\ket{1}\leftrightarrow \ket{2}$  & - & - &  - & $\dots$ & - & - \\
$\psi_3 \geq \phi_3$ & $\ket{1}\leftrightarrow \ket{3}$  & $\ket{2}\leftrightarrow \ket{3}$ & - & - & $\dots$ & - & - \\
$\psi_4 \geq \phi_4$ & $\ket{1}\leftrightarrow \ket{4}$  & $\ket{2}\leftrightarrow \ket{4}$ & $\ket{3}\leftrightarrow \ket{4}$ & - & $\dots$ & - & - \\
$\psi_5 \geq \phi_5$ & $\ket{1}\leftrightarrow \ket{5}$  & $\ket{2}\leftrightarrow \ket{5}$ & $\ket{3}\leftrightarrow \ket{5}$ & $\ket{4}\leftrightarrow \ket{5}$ & $\dots$ & - & - \\
$\vdots$ & $\vdots$ & $\vdots$ & $\vdots$ & $\vdots$ & $\vdots$ & $\vdots$ & $\vdots$  \\
$\psi_{d-2} \geq \phi_{d-2}$ & $\ket{1}\leftrightarrow \ket{d-2}$  & $\ket{2}\leftrightarrow \ket{d-2}$ & $\ket{3}\leftrightarrow \ket{d-2}$ & $\ket{4}\leftrightarrow \ket{d-2}$ & $\dots$ & - & -  \\
$\psi_{d-1} \geq \phi_{d-1}$ & $\ket{1}\leftrightarrow \ket{d-1}$  & $\ket{2}\leftrightarrow \ket{d-1}$ & $\ket{3}\leftrightarrow \ket{d-1}$ & $\ket{4}\leftrightarrow \ket{d-1}$ & $\dots$ & $\ket{d-2}\leftrightarrow \ket{d-1}$ & -\\
$\psi_d \geq \phi_d$ & $\ket{1}\leftrightarrow \ket{d}$ & $\ket{2}\leftrightarrow \ket{d}$ & $\ket{3}\leftrightarrow \ket{d}$ & $\ket{4}\leftrightarrow \ket{d}$ & $\dots$ & $\ket{d-2}\leftrightarrow \ket{d}$ & $\ket{d-1}\leftrightarrow \ket{d}$ \\  [1ex] 
\end{tabular}
\end{ruledtabular}
\label{table-n}
\end{table*}
One of the main advantages of this protocol is that it provides a single map for $\ket{\psi} \xrightarrow{\text{ico}} \ket{\phi}$ (ico stands for incoherently). Let us assume that there are Kraus operators of the form
\begin{eqnarray}\label{Kraus-n}
K_{s}^i=U_{s}^i\left(\sqrt{p_s^i}\sum_{j=1}^{d}\frac{c_{sij}}{\psi_j}\ket{j}\bra{j}\right)=U_{s}^i M_{s}^i,
\end{eqnarray}
which constitute a single map $\Phi_s$ for the transformation $\ket{\psi}\bra{\psi} \rightarrow \ket{\phi}\bra{\phi}$ such that
\begin{eqnarray}\label{map-n}
\Phi_s[\rho_{\psi}]=\sum_{i=1}^{d}K_s^i \rho_{\psi} {K_s^i}^{\dag}=\rho_{\phi},
\end{eqnarray}
where $\sum_{i=1}^{d} {K_s^i}^{\dag}K_s^i=I_d$ and $p_s^i={\text{Tr}}[K_s^i \rho_{\psi} {K_s^i}^{\dag}]$. It is obvious that quantum operation  $\Phi_s$ is an incoherent operation whose Kraus operators are also incoherent.
In Eq. \eqref{Kraus-n} $U_s^i$ is the $i$th element of the set of permutations $U_s$ where $s=1,2,\dots,n$.
A partial detailed exposition for $s$ follows below. We express the coefficients $c_{sij}$ in compact form as the elements of the matrix $c_s$, i.e., $c_{sij}$ is the $(ij)$th element of the $d\times d$ matrix $c_s$ where
\begin{eqnarray}\label{unitarytrss}
U_s^i (c_{si1}, c_{si2}, \dots, c_{sid})^{T}=(\phi_1, \phi_2, \dots, \phi_d)^{T}.
\end{eqnarray}
We may also interpret the deterministic transformation \eqref{map-n} as a $d$-outcome positive-operator valued measure
(POVM) with measurement operators
\begin{eqnarray}\label{POVM}
M_s^i=\sqrt{p_s^i}\sum_{j=1}^{d}
\frac{c_{sij}}{\psi_j}\ket{j}\bra{j},
\end{eqnarray}
satisfying $\sum_{i=1}^{d}{M_s^i}^{\dag}M_s^i=I_d$. The POVM measurement yields
\begin{equation}
M_s^i\ket{\psi}\rightarrow\ket{\varphi_s^i}=\sum_{j=1}^{d}c_{sij}\ket{j} \quad (i=1, \dots, d),
\end{equation}
with probabilities $p_s^i=\bra{\psi}{M_s^i}^{\dag}M_s^i\ket{\psi}$ where $\ket{\varphi_s^1}$ is equal to $\ket{\phi}$, and $\ket{\varphi_s^i}$ ($i=2,3,...,d$) is equal to $\ket{\phi}$ up to permutation of the basis $\ket{j}$, i.e., $U_s^i\ket{\varphi_s^i}=\ket{\phi}$.
The crucial point of the protocol is to find the correct set of $d$ states $\ket{\varphi_s^i}$ obtained as a result of the measurement.
In general, there are many states which are equal to the target state $\ket{\phi}$ up to permutations of the basis, and there are too many possibilities to choose sets with $d$ elements.
However, there are only few sets which satisfy the conditions on the measurement, positivity of $p_s^i$ and $\sum_{i}p_s^i=1$. For a complete set of a given case we need $n$ sets of permutations (where $s=1,2,\dots,n$) which fulfill the positivity of $p_s^i$. We encode the correct sets in the $d$ unitary operations (permutations $U_s^1, U_s^2, \dots, U_s^d$) given in the solutions. These correct sets of states and the permutation sets depend on the relations between the coefficients of the initial and final states.
Hence, finding the set of permutations (finding $c_{sij}$) is a highly nontrivial problem, and the problem becomes exponentially difficult as the dimension becomes greater.
However, we are able to propose a complete solution (a permutation-based protocol) for $d$-level systems.
We first identify the set of permutations (SP), $U_s$, and using these we obtain both probabilities and a set of Kraus operators.
The condition $\sum_{i=1}^{d}{K_s^i}^{\dag}K_s^i=I_d$  implies that the following $s$ sets of $d$ equations
\begin{eqnarray}\begin{aligned}\label{probabilities}
&\sum_{i=1}^{d}p_s^i{c^2_{sij}}=\psi^2_j \quad (j=1,2,\dots,d),& \\ &\Big(\sum_{i=1}^{d}p_s^i=1, \quad  p_s^i\geq 0\Big),&
\end{aligned}\end{eqnarray}
should be satisfied. Here, the coefficients $c_{sij}$ are given by
\begin{equation}\label{csij}
(c_{si1}, c_{si2}, \dots, c_{sid})^{T}={U_s^i}^{\dag}(\phi_1, \phi_2, \dots, \phi_d)^{T}.
\end{equation}
Hence, given the SPs (we propose a method how to find them), the problem is reduced to solving $d$ linear equations with $d$ unknowns $p_s^i$ ($i=1,2,\dots,d$), and the solutions of Eq. \eqref{probabilities} determine the solutions for the Kraus operators $K_s^i$ in Eq. \eqref{Kraus-n} which make the transformation $\rho_{\psi} \rightarrow \rho_{\phi}$.
Succinctly, the problem is to find the correct SPs in Eq. \eqref{unitarytrss}.

Although the relations $\psi_1 \leq \phi_1$ and $\psi_d \geq \phi_d$ follow from the majorization, there are $2^{d-2}$
($d>2$) possible relations for the other coefficients of initial and final coherent states.
The first step, the crux of our method, is to construct a table (see Table \ref{table-n}) which contains all possible permutations for a case (either $\psi_k\leq\phi_k$ or $\psi_k\geq\phi_k$ where $k=2,3,...,d-1$) of $d$-level systems.
In a table there will be a certain number $\zeta$ of permutations where $(d-1)\leq\zeta\leq \lfloor d/2\rfloor \lceil d/2 \rceil$. The $(d-1)$ element combinations  of these permutations constitute SPs together with $I_d$ ($d \times d$ identity transformation). A single SP $(s=1)$ is sufficient if $\zeta=d-1$, and there are $d-1$ single SP for $d$-level systems. On the other hand, if $\zeta > d-1$, then the given case splits into subcases, and therefore, for a complete set we need $n>1$ ($s=1,2,\dots,n$) SPs for the corresponding case, i.e., number of subcases is $n$. For different coefficients satisfying different relations, we need a different SP and set of Kraus operators. We label each set by the index ``$s$'' which refers to a solution for a given relation (subcase relation) between coefficients. Now, before we begin to explain our protocol, we define
\begin{eqnarray}\begin{aligned}
\phi_x^2+\phi_{x+1}^2+ \dots +\phi_{k}^2-\psi_x^2-\psi_{x+1}^2- \dots -\psi_{k}^2 & \equiv \alpha_{x(x+1) \dots k}, \quad \\
\psi_l^2+\psi_{l+1}^2+ \dots +\psi_{y}^2-\phi_l^2-\phi_{l+1}^2- \dots -\phi_{y}^2 & \equiv \beta_{l(l+1) \dots y}, \quad \\
\phi_x^2-\phi_y^2 & \equiv \gamma_{xy},
\end{aligned}\end{eqnarray}
where $y > x$ and $\gamma_{xy}>0$. Given the correct set of permutations, the solutions of Eq. \eqref{probabilities} for probability, corresponding to the permutation $\ket{x}\leftrightarrow\ket{y}$, turn out to be either
\begin{eqnarray}
p=\frac{\alpha_{x(x+1)(x+2)\dots k}}{\gamma_{xy}},
\end{eqnarray}
where $k\in [x,y-1]$, or
\begin{eqnarray}
p=\frac{\beta_{l(l+1)(l+2)\dots y}}{\gamma_{xy}},
\end{eqnarray}
where $l\in [x+1,y]$.
In our protocol, the set of permutations is obtained as follows:
We first construct  a table using the relations between the  coefficients of the initial (source) and the final (target) states. Then,

\begin{enumerate}[(i)]

\item All sets of permutations contain the identity transformation,
      $U_s^1=I_d$. The measurement (where the Kraus operator is  $K_s^1$)
      probability corresponding to this permutation is found to be $p_s^1=1-\sum_{i=2}^{d} p_s^i$.

\item The permutation $\ket{v}\leftrightarrow \ket{m}$ exists in all the
      sets of permutations if $\ket{v}\leftrightarrow \ket{m}$
      is the only permutation in a column of the resulting table. The measurement probability corresponding to this permutation is found to be $(\phi_v^2-\psi_v^2)/(\phi_v^2-\phi_m^2)\equiv {\alpha_v}/{\gamma_{vm}}$.

\item The permutation $\ket{h}\leftrightarrow \ket{k}$ exists in all the
      sets of permutations if $\ket{h}\leftrightarrow \ket{k}$
      is the only permutation in a row of the resulting table. The measurement probability corresponding to this permutation is found to be $(\psi_k^2-\phi_k^2)/(\phi_h^2-\phi_k^2)\equiv {\beta_k}/{\gamma_{hk}}$.

\item If any permutation $\ket{u}\leftrightarrow \ket{u+1}$ appears in
      the table then it must be an element of all the sets of
      permutations. The measurement probability corresponding to this permutation is either $(\psi_{u+1}^2-\phi_{u+1}^2)/(\phi_u^2-\phi_{u+1}^2)\equiv {\beta_{u+1}}/{\gamma_{u(u+1)}}$ or $(\phi_u^2-\psi_u^2)/(\phi_u^2-\phi_{u+1}^2)\equiv {\alpha_{u}}/{\gamma_{u(u+1)}}$.

\item All sets of permutations contain the permutation
      $\ket{1}\leftrightarrow \ket{d}$ ($U_s^{r}=\ket{1}\leftrightarrow
      \ket{d}\equiv\ket{d}\bra{1}+\sum_{i=2}^{d-1}\ket{i}\bra{i}
      +\ket{1}\bra{d}$).

\end{enumerate}
After applying the above steps as a first stage of our protocol, we get the set(s) of permutations of the form
\begin{eqnarray}\begin{aligned}\label{unitarytrs}
U_s&=\{U_s^1, U_s^2, \dots, U_s^r, U_s^{r+1}, \dots, U_s^d\} \\
&=\{I_d, \ket{v}\leftrightarrow \ket{m}, \ket{h}\leftrightarrow \ket{k}, \ket{u}\leftrightarrow \ket{u+1}, \\ & \ \ \quad \dots, U_s^{r-1}, \ket{1}\leftrightarrow \ket{d}, U_s^{r+1}, \dots, U_s^d\}.
\end{aligned}\end{eqnarray}
We note that a table may contain more than one of permutations $\ket{v}\leftrightarrow \ket{m}$, $\ket{h}\leftrightarrow \ket{k}$ and $\ket{u}\leftrightarrow \ket{u+1}$. Therefore, $r$ permutations are found after five steps described by (i)-(v). We still need a certain number, $d-r$, of permutations to complete the set $U_s$. In order to be able to explain how the remaining permutations, $\{U_s^{r+1}, U_s^{r+2}, \dots, U_s^d\}$, are chosen, we give illustrative examples with pictorial representations.

Using the pictorial representation of the table of permutations, we  observe that Eq. \eqref{probabilities} has a solution with positive probabilities for all SPs with no intersection of permutations. Each nonintersecting set of permutations corresponds to a solution for an additional relation (subcase) between the coefficients of the states. More importantly, the collection of the solutions of Eq. \eqref{probabilities} for all subcases forms the complete solution and the collection of the SPs with no intersection in the pictorial representation forms what we call the ``complete set.''

Hence, the problem of obtaining the correct sets of permutations ($U_s$) is reduced to the problem of obtaining sets of nonintersecting permutations in the pictorial representations.
It is also obvious that the permutations described by (i)$-$(v) ($I_d$, $\ket{v}\leftrightarrow \ket{m}$, $\ket{h}\leftrightarrow \ket{k}$, $\ket{u}\leftrightarrow \ket{u+1}$, $\ket{1}\leftrightarrow \ket{d}$) do intersect each other and any other permutations in the pictorial representation of any table.



\subsubsection{Examples for protocol I}\label{coherent-examples}


We can best understand protocol I by explicitly examining some examples; we give two illustrative examples.
As a first example, consider the case $\psi_2 \leq \phi_2$ and $\psi_3 \geq \phi_3$ for $d=4$ where $\psi_1 \leq \phi_1$ and $\psi_4 \geq \phi_4$ follow from the majorization condition. In the beginning we construct Table \ref{table-d4} which contains all possible permutations for this case.
\begin{table}[!htbp]
\caption{All possible permutations for the case $\psi_1 \leq \phi_1$, $\psi_2 \leq \phi_2$, $\psi_3 \geq \phi_3$ and $\psi_4 \geq \phi_4$ of $d=4$.}
\label{table-d4}
\centering
\begin{ruledtabular}
\begin{tabular}{c c c c}
         & $\psi_1 \leq \phi_1$ & $\psi_2 \leq \phi_2$   \\ [0.5ex] \hline
$\psi_3 \geq \phi_3$ & $\ket{1}\leftrightarrow \ket{3}$ & $\ket{2}\leftrightarrow \ket{3}$   \\
$\psi_4 \geq \phi_4$ & $\ket{1}\leftrightarrow \ket{4}$ & $\ket{2}\leftrightarrow \ket{4}$   \\
\end{tabular}
\end{ruledtabular}
\end{table}
There are four permutations in Table \ref{table-d4}. Figure \ref{figure-n4all} shows the pictorial representation of these. The three-element combinations of these four permutations constitute the SPs together with identity transformation $I_4$, where Fig. \ref{figure-n4} provides the pictorial representations of all possible three-element combinations.
\begin{subfigures}
	\begin{figure}[!htbp]
		\centering
	 \includegraphics[width=0.09\textwidth]{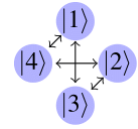}
		\caption{Pictorial representation of four permutations given in Table \ref{table-d4} for the case $\psi_2 \leq \phi_2$ and $\psi_3 \geq \phi_3$ of four-level states.}\label{figure-n4all}
	\end{figure}
	\begin{figure}[!htbp]
		\centering
	\includegraphics[width=0.45\textwidth]{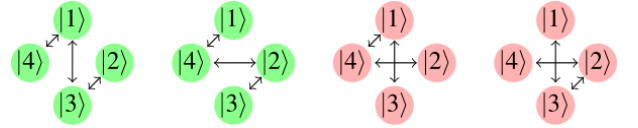}
		\caption{Pictorial representations of all possible three-element combinations of the permutations given in Table \ref{table-d4}. Here, the first two three-element combinations of the permutations (in green) are sufficient to constitute a complete set (together with $I_4$). However, last two (in red) are not usable where these sets give negative probabilities (which is unphysical).}
		\label{figure-n4}
	\end{figure}
\end{subfigures}

According to our protocol, any set(s) of permutations contain the identity transformation, $I_4$. We have also proposed that if any permutation $\ket{u}\leftrightarrow \ket{u+1}$ appears in the table then it must be an element of all SPs, $\ket{2}\leftrightarrow \ket{3}$ belongs to this category. The permutation  $\ket{1}\leftrightarrow \ket{4}$ must also be an element of all SPs. Thus, the SPs have been obtained such that
\begin{eqnarray}
U_s=\Big\{I_4, \ket{2}\leftrightarrow \ket{3}, \ket{1}\leftrightarrow \ket{4},  U_s^4\Big\}.
\end{eqnarray}
The remaining permutation, $U_s^4$, is chosen among the permutations $\ket{1}\leftrightarrow \ket{3}$ and $\ket{2}\leftrightarrow \ket{4}$. We obtained that $U_1^4=\ket{1}\leftrightarrow \ket{3}$ and $U_2^4=\ket{2}\leftrightarrow \ket{4}$. The SPs were identified, and using these we can obtain the two sets of Kraus operators. The Kraus operators are given by
\begin{eqnarray}
K_s^i=U_s^i\Big(\sqrt{p_s^i}\sum_{j=1}^{4}\frac{c_{sij}}{\psi_j}\ket{j}\bra{j}\Big) \quad (s=1,2),
\end{eqnarray}
where $\sum_{i=1}^{4} {K_s^i}^{\dag}K_s^i=I_4$ and $p_s^i={\text{Tr}}[K_s^i \rho_{\psi} {K_s^i}^{\dag}]$. We express the coefficients $c_{sij}$ in compact form as the elements of the matrix $c_s$ , i.e., $c_{sij}$ is the $(ij)$th element of the matrix $c_s$ where $U_s^i (c_{si1}, c_{si2}, c_{si3}, c_{si4})^{T}=(\phi_1, \phi_2, \phi_3, \phi_4)^{T}$. The matrix $c_s$ is given by
\begin{eqnarray}\begin{aligned}
c_s
=\left(\begin{array}{cccccc}  \phi_1 & \phi_2 & \phi_3 & \phi_4 \\ \phi_1 & \phi_3 & \phi_2 & \phi_4 \\ \phi_4 & \phi_2 & \phi_3 & \phi_1 \\  c_{s41} &  c_{s42} &  c_{s43} &  c_{s44} \end{array}\right).
\end{aligned}\end{eqnarray}
The condition $\sum_{i=1}^{4} {K_s^i}^{\dag}K_s^i=I_4$ implies that the following two sets of four linear equations
\begin{eqnarray}\label{xij}
\sum_{i=1}^{4} {p_s^i} {c^2_{sij}}=\psi^2_j, \quad (j=1,2,3,4), \quad (s=1,2),
\end{eqnarray}
should be satisfied with the constraint $p_s^i\geq 0$. Solutions of the linear equations $\sum_{i=1}^{4} {p_1^i} {c^2_{1ij}}=\psi^2_j$ gives the probabilities of the first set  where the
matrix $c_1$ is given by (follows from $U_1=\{I_4, \ket{2}\leftrightarrow \ket{3}, \ket{1}\leftrightarrow \ket{4}, \ket{1}\leftrightarrow \ket{3}\}$)
\begin{eqnarray}\label{matrixn4-1}
c_1=\left(\begin{array}{cccccc}  \phi_1 & \phi_2 & \phi_3 & \phi_4 \\ \phi_1 & \phi_3 & \phi_2 & \phi_4 \\ \phi_4 & \phi_2 & \phi_3 & \phi_1 \\  \phi_3 & \phi_2 & \phi_1 & \phi_4 \end{array}\right).
\end{eqnarray}
Thus, probabilities of the first set are found to be
\begin{eqnarray}\begin{aligned}
p_1^1=1-\sum_{i=2}^{4} p_1^i, \quad
p_1^2=\frac{\alpha_2}{\gamma_{23}}
, \quad
p_1^3=\frac{\beta_4}{\gamma_{14}}
, \quad
p_1^4=\frac{\beta_{23}}{\gamma_{13}}
, \quad
\end{aligned}\end{eqnarray}
where $\beta_{23}\geq 0$, i.e, $\psi_2^2+\psi_3^2 \geq \phi_2^2+\phi_3^2$. Similarly, solutions of the linear equations $\sum_{i=1}^{4} {p_2^i} {c^2_{2ij}}=\psi^2_j$ gives the probabilities of the second set where the matrix $c_2$ is given by (follows from $U_2=\{I_4, \ket{2}\leftrightarrow \ket{3}, \ket{1}\leftrightarrow \ket{4}, \ket{2}\leftrightarrow \ket{4}\}$)
\begin{eqnarray}\label{matrixn4-1}
c_2=\left(\begin{array}{cccccc}  \phi_1 & \phi_2 & \phi_3 & \phi_4 \\ \phi_1 & \phi_3 & \phi_2 & \phi_4 \\ \phi_4 & \phi_2 & \phi_3 & \phi_1 \\  \phi_1 & \phi_4 & \phi_3 & \phi_2 \end{array}\right).
\end{eqnarray}
Then, the probabilities of the second set are found to be
\begin{eqnarray}\begin{aligned}
p_2^1=1-\sum_{i=2}^{4} p_1^i, \quad
p_2^2=\frac{\beta_3}{\gamma_{23}} 
, \quad
p_2^3=\frac{\alpha_1}{\gamma_{14}}
, \quad
p_2^4=\frac{\alpha_{23}}{\gamma_{24}} 
, \quad
\end{aligned}\end{eqnarray}
where $\alpha_{23}\geq 0$, i.e., $\psi_2^2+\psi_3^2 \leq \phi_2^2+\phi_3^2$. This case, $\psi_1 \leq \phi_1$, $\psi_2 \leq \phi_2$, $\psi_3 \geq \phi_3$ and $\psi_4 \geq \phi_4$ for $d=4$, splits into two subcases. If the subcase relation is $\psi_2^2+\psi_3^2 \geq \phi_2^2+\phi_3^2$ ($\beta_{23}\geq 0$) then the first SP, $U_1,$ must be used. If the subcase relation is $\psi_2^2+\psi_3^2 \leq \phi_2^2+\phi_3^2$ ($\alpha_{23}\geq 0$) then the second SP, $U_2$, must be used. These two SPs constitute the complete set and give the complete solution of the problem, i.e., the probabilities and Kraus operators are found for all possible relations among the coefficients for the given case $\psi_1 \leq \phi_1$, $\psi_2 \leq \phi_2$, $\psi_3 \geq \phi_3$ and $\psi_4 \geq \phi_4$ of $d=4$.


As a second example, consider the case $\psi_a \leq \phi_a$ ($a=1,4,5,7,8$) and $\psi_b \geq \phi_b$ ($b=2,3,6,9$)  for $d=9$.
\begin{table}[!htbp]
\caption{All possible permutations for the case  $\psi_a \leq \phi_a$ ($a=1,4,5,7,8$) and $\psi_b \geq \phi_b$ ($b=2,3,6,9$) of $d=9$.}
\label{table-n10}
\centering
\begin{ruledtabular}
\begin{tabular}{c c c c c c c c c c}
         & $\psi_1 \leq \phi_1$ & $\psi_4 \leq \phi_4$ & $\psi_5 \leq \phi_5$ &  $\psi_7 \leq \phi_7$ & $\psi_8 \leq \phi_8$  \\ [0.5ex] \hline
$\psi_2 \geq \phi_2$ & $\ket{1}\leftrightarrow \ket{2}$ & - & - & - &- \\
$\psi_3 \geq \phi_3$ & $\ket{1}\leftrightarrow \ket{3}$ & - & - & - &- \\
$\psi_6 \geq \phi_6$ & $\ket{1}\leftrightarrow \ket{6}$ & $\ket{4}\leftrightarrow \ket{6}$ & $\ket{5}\leftrightarrow \ket{6}$ & - & - \\
$\psi_{9} \geq \phi_{9}$ & $\ket{1}\leftrightarrow \ket{9}$ & $\ket{4}\leftrightarrow \ket{9}$ & $\ket{5}\leftrightarrow \ket{9}$ & $\ket{7}\leftrightarrow \ket{9}$ & $\ket{8}\leftrightarrow \ket{9}$\\
\end{tabular}
\end{ruledtabular}
\end{table}
In Table \ref{table-n10}, there are ten permutations, some eight-element combinations of which, together with the identity transformation, constitute the permutations $U_s^i$ ($i=1,\dots,9$).
In our protocol, the SPs are obtained in five steps described in (i)$-$(v) plus an additional step (using nonintersecting permutations in the pictorial representations). Let us consider each step in turn.
First, all SPs contain the identity transformation; $U_s^1=I_{9}$.
Second, the permutation $\ket{v}\leftrightarrow \ket{m}$ exists in all the SPs if $\ket{v}\leftrightarrow \ket{m}$ is the only permutation in any column of Table \ref{table-n10}; $U_s^2=\ket{7}\leftrightarrow \ket{9}$ and $U_s^3=\ket{8}\leftrightarrow \ket{9}$.
Third, the permutation $\ket{h}\leftrightarrow \ket{k}$ exists in all the SPs if $\ket{h}\leftrightarrow \ket{k}$ is the only permutation in any row of Table \ref{table-n10}; $U_s^4=\ket{1}\leftrightarrow \ket{2}$ and $U_s^5=\ket{1}\leftrightarrow \ket{3}$.
Fourth, if any permutation $\ket{u}\leftrightarrow \ket{u+1}$ appears in the table then it must be an element of all the SPs; $U_s^6=\ket{5}\leftrightarrow \ket{6}$.
Fifth, all SPs contain the permutation $\ket{1}\leftrightarrow \ket{d}$; $U_s^7=\ket{1}\leftrightarrow \ket{9}$. Thus, after five steps, the point we have reached is that
\begin{eqnarray}\begin{aligned}\label{SPford9}
U_s=\big\{&I_{9}, \ket{7}\leftrightarrow \ket{9}, \ket{8}\leftrightarrow \ket{9}, \ket{1}\leftrightarrow \ket{2}, \ket{1}\leftrightarrow \ket{3},  \\ & \ket{5}\leftrightarrow \ket{6},\ket{1}\leftrightarrow \ket{9}, U_s^8,  U_s^9\big\}.
\end{aligned}\end{eqnarray}
The SPs given by Eq. \eqref{SPford9} satisfies our interesting result$-$any pairs of permutations in a SP do not intersect each other in the pictorial representation as seen in Fig. \ref{figure-n10}.
\begin{subfigures}
	\begin{figure}[!htbp]
		\centering
		\includegraphics[width=0.2\textwidth]{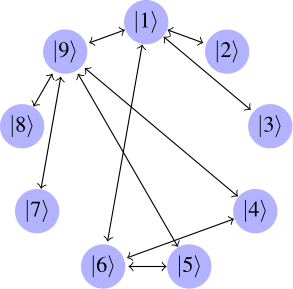}
		\caption{Pictorial representation of ten permutations given in Table \ref{table-n10} for the case  $\psi_a \leq \phi_a$ ($a=1,4,5,7,8$) and $\psi_b \geq \phi_b$ ($b=2,3,6,9$) of $d=9$. The permutations $\ket{7}\leftrightarrow \ket{9}$, $\ket{8}\leftrightarrow \ket{9}$, $\ket{1}\leftrightarrow \ket{2}$, $\ket{1}\leftrightarrow \ket{3}$, $\ket{5}\leftrightarrow \ket{6}$, and $\ket{1}\leftrightarrow \ket{9}$ do not intersect each other and any other permutations.}
		\label{figure-n10}
	\end{figure}
	\begin{figure}[!htbp]
		\centering
		\includegraphics[width=0.45\textwidth]{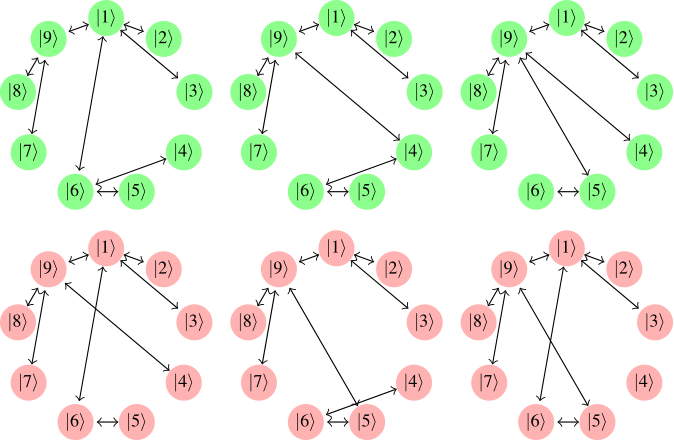}
		\caption{Pictorial representations of the SPs (also $I_9$ is an element of all SPs) for the case  $\psi_a \leq \phi_a$ ($a=1,4,5,7,8$) and $\psi_b \geq \phi_b$ ($b=2,3,6,9$) of $d=9$. Here, the first three eight-element combinations of the permutations (in green) are sufficient to constitute a complete set.
		 However, last three eight-element combinations of the permutations (in red) are not usable while these give negative probabilities.}
		\label{figure-d9}
	\end{figure}
\end{subfigures}
We still need two more permutations, $U_s^8$ and $U_s^{9}$, and these will be selected from among four remaining permutations, $\ket{1}\leftrightarrow \ket{6}$, $\ket{4}\leftrightarrow \ket{6}$, $\ket{4}\leftrightarrow \ket{9}$ and $\ket{5}\leftrightarrow \ket{9}$. The final step is to determine the correct remaining permutations for SPs to constitute a complete set. The result we have obtained is that three SPs ($s=1,2,3$) among six combinations are sufficient [see Fig. \ref{figure-d9}] to constitute a complete set, i.e., for a complete solution, three SPs are required which equally means that three subcases exist for the given case.
Thus, two remaining permutations for these three SPs are found to be
\begin{eqnarray}\begin{aligned}\label{SPford9remain}
\big( U_1^8, U_1^9\big)&=\big(\ket{1}\leftrightarrow \ket{6},  \ket{4}\leftrightarrow \ket{6}\big), \\
\big(U_2^8, U_2^9\big)&=\big(\ket{4}\leftrightarrow \ket{9},  \ket{4}\leftrightarrow \ket{6}\big), \\
\big(U_3^8, U_3^9\big)&=\big(\ket{4}\leftrightarrow \ket{9},  \ket{5}\leftrightarrow \ket{9}\big).
\end{aligned}\end{eqnarray}
The SPs were identified [combine Eq. \eqref{SPford9} and Eq. \eqref{SPford9remain}], and using these we obtain the sets of Kraus operators. The Kraus operators are given by
\begin{eqnarray}
K_s^i=U_s^i\Big(\sqrt{p_s^i}\sum_{j=1}^{9}\frac{c_{sij}}{\psi_j}\ket{j}\bra{j}\Big) \quad (s=1,2,3),
\end{eqnarray}
where $(c_{si1}, c_{si2}, c_{si3}, \dots, c_{si9})^{T}=
{(U_s^i)}^{\dag} (\phi_1, \phi_2, \phi_3, \dots, \phi_{9})^{T}$ and $p_s^i={\text{Tr}}[K_s^i \rho_{\psi} {K_s^i}^{\dag}]$. The matrix $c_s$ is given by
\begin{eqnarray}\begin{aligned}
c_s
=\left(\begin{array}{ccccccccc}  \phi_1 & \phi_2 & \phi_3 & \phi_4 & \phi_5 & \phi_6 & \phi_7 & \phi_8 & \phi_9 \\ \phi_1 & \phi_2 & \phi_3 & \phi_4 & \phi_5 & \phi_6 & \phi_9 & \phi_8 & \phi_7 \\ \phi_1 & \phi_2 & \phi_3 & \phi_4 & \phi_5 & \phi_6 & \phi_7 & \phi_9 & \phi_8 \\ \phi_2 & \phi_1 & \phi_3 & \phi_4 & \phi_5 & \phi_6 & \phi_7 & \phi_8 & \phi_9 \\ \phi_3 & \phi_2 & \phi_1 & \phi_4 & \phi_5 & \phi_6 & \phi_7 & \phi_8 & \phi_9 \\ \phi_1 & \phi_2 & \phi_3 & \phi_4 & \phi_6 & \phi_5 & \phi_7 & \phi_8 & \phi_9 \\  \phi_9 & \phi_2 & \phi_3 & \phi_4 & \phi_5 & \phi_6 & \phi_7 & \phi_8 & \phi_1 \\  c_{s81} &  c_{s82} &  c_{s83} &  c_{s84} & c_{s85} &  c_{s86} &  c_{s87} &  c_{s88} & c_{s89} \\  c_{s91} &  c_{s92} &  c_{s93} &  c_{s94} & c_{s95} &  c_{s96} &  c_{s97} &  c_{s98} & c_{s99} \end{array}\right).  \; \; \; \quad
\end{aligned}\end{eqnarray}
The condition $\sum_{i=1}^{9} {K_s^i}^{\dag}K_s^i=I_{9}$ implies the following linear equations
\begin{eqnarray}\label{csij9}
\sum_{i=1}^{9} {p_s^i} {c^2_{sij}}=\psi^2_j, \quad  p_s^i\geq 0, \quad (j=1,2,\dots,9),
\end{eqnarray}
whose solutions for $p_s^i$ give the probabilities for each set. The probabilities corresponding to the first SP $U_1$ (solutions of the linear equations $\sum_{i=1}^{9} {p_1^i} {c^2_{1ij}}=\psi^2_j$) are found to be
\begin{eqnarray}\begin{aligned}
p_1^1&=1-\sum_{i=2}^{9} p_s^i, \quad
p_1^2=\frac{\alpha_7}{\gamma_{79}}, \quad
p_1^3=\frac{\alpha_8}{\gamma_{89}}, \quad
p_1^4=\frac{\beta_2}{\gamma_{12}}, \\
p_1^5&=\frac{\beta_3}{\gamma_{13}} \quad
p_1^6=\frac{\alpha_5}{\gamma_{56}}, \quad
p_1^7=\frac{\beta_{789}}{\gamma_{19}}, \\
p_1^8&=\frac{\beta_{456}}{\gamma_{16}}, \quad
p_1^9=\frac{\alpha_4}{\gamma_{46}},
\end{aligned}\end{eqnarray}
where $\beta_{456}\geq 0$. The probabilities corresponding to the second SP $U_2$ (solutions of the linear equations $\sum_{i=1}^{9} {p_2^i} {c^2_{2ij}}=\psi^2_j$) are found to be
\begin{eqnarray}\begin{aligned}
p_2^1&=1-\sum_{i=2}^{9} p_s^i, \quad
p_2^2=\frac{\alpha_7}{\gamma_{79}}, \quad
p_2^3=\frac{\alpha_8}{\gamma_{89}}, \quad
p_2^4=\frac{\beta_2}{\gamma_{12}}, \\
p_2^5&=\frac{\beta_3}{\gamma_{13}} \quad
p_2^6=\frac{\alpha_5}{\gamma_{56}}, \quad
p_2^7=\frac{\alpha_{123}}{\gamma_{19}}, \\
p_2^8&=\frac{\alpha_{456}}{\gamma_{49}}, \quad
p_2^9=\frac{\beta_{56}}{\gamma_{46}},
\end{aligned}\end{eqnarray}
where $\alpha_{456}\geq 0$ and $\beta_{56}\geq 0$. The probabilities corresponding to the third SP $U_3$ (solutions of the linear equations $\sum_{i=1}^{9} {p_3^i} {c^2_{3ij}}=\psi^2_j$) are found to be
\begin{eqnarray}\begin{aligned}
p_3^1&=1-\sum_{i=2}^{9} p_s^i, \quad
p_3^2=\frac{\alpha_7}{\gamma_{79}}, \quad
p_3^3=\frac{\alpha_8}{\gamma_{89}}, \quad
p_3^4=\frac{\beta_2}{\gamma_{12}}, \\
p_3^5&=\frac{\beta_3}{\gamma_{13}} \quad
p_3^6=\frac{\beta_6}{\gamma_{56}}, \quad
p_3^7=\frac{\alpha_{123}}{\gamma_{19}}, \\
p_3^8&=\frac{\alpha_4}{\gamma_{49}}, \quad
p_3^9=\frac{\alpha_{56}}{\gamma_{59}},
\end{aligned}\end{eqnarray}
where $\alpha_{56}\geq 0$. We note that the probabilities $p_s^1$, $p_s^2$, $p_s^3$, $p_s^4$ and $p_s^5$ turn out to be of the form
\begin{eqnarray}\begin{aligned}\label{knownprob}
p_s^1&=1-\sum_{i=2}^{9} p_s^i, \quad
p_s^2=\frac{\alpha_7}{\gamma_{79}}, \quad
p_s^3=\frac{\alpha_8}{\gamma_{89}}, \\
p_s^4&=\frac{\beta_2}{\gamma_{12}}, \quad
p_s^5=\frac{\beta_3}{\gamma_{13}} \quad (s=1,2,3),
\end{aligned}\end{eqnarray}
as we stated in (i), (ii) and (iii). We obtained that there are three SPs for this given case. Then, if the subcase is $\psi^2_{4}+\psi^2_{5}+\psi^2_{6} \geq \phi^2_{4}+\phi^2_{5}+\phi^2_{6}$ ($\beta_{456}\geq 0$) then the SP $U_1$ will be used.  If the  subcase is $\psi^2_{4}+\psi^2_{5}+\psi^2_{6} \leq \phi^2_{4}+\phi^2_{5}+\phi^2_{6}$ and $\psi^2_{5}+\psi^2_{6} \geq \phi^2_{5}+\phi^2_{6}$ ($\alpha_{456}\geq 0$ and  $\beta_{56}\geq 0$) then the SP $U_2$ will be used. If the subcase is $\psi^2_{5}+\psi^2_{6} \leq \phi^2_{5}+\phi^2_{6}$ ($\alpha_{56}\geq 0$) then the SP $U_3$ will be used. Thus, these three SPs ($s=1,2,3$) form a complete set for the given case$-$using one of these three SPs one can transform nine-level coherent states, whose coefficients have the relations $\psi_a \leq \phi_a$ ($a=1,4,5,7,8$) and $\psi_b \geq \phi_b$ ($b=2,3,6,9$), via incoherent operations.

\begin{table*}[!htbp]
	\caption{All possible permutations for the case $\psi_1 \leq \phi_1$, $\psi_2 \geq \phi_2$,  $\psi_3 \geq \phi_3$, $\dots$, $\psi_{k-1} \geq \phi_{k-1}$, $\psi_{k} \geq \phi_{k}$, $\psi_{k+1} \leq \phi_{k+1}$, $\psi_{k+2} \leq \phi_{k+2}$, $\dots$, $\psi_{d-1} \leq \phi_{d-1}$, $\psi_d \geq \phi_d$. There are $d-1$ permutations for this case, and a single set of permutations is sufficient for complete solution.}
	\label{gen} 
	\centering 
	\begin{ruledtabular}
		\begin{tabular}{c c c c c c c c}
			& $\psi_1 \leq \phi_1$ & $\psi_{k+1} \leq \phi_{k+1}$ & $\psi_{k+2} \leq \phi_{k+2}$ & $\psi_{k+3} \leq \phi_{k+3}$ & \dots & $\psi_{d-2} \leq \phi_{d-2}$ & $\psi_{d-1} \leq \phi_{d-1}$ \\ [0.5ex] \hline
			$\psi_2 \geq \phi_2$ & $\ket{1}\leftrightarrow \ket{2}$ & - & - & - & \dots & - & - \\
			$\psi_3 \geq \phi_3$ & $\ket{1}\leftrightarrow \ket{3}$ & - & - & - & \dots & - & - \\
			$\vdots$ & $\vdots$  & $\vdots$ & $\vdots$  & $\vdots$ & $\vdots$ & $\vdots$ & $\vdots$ \\
			$\psi_{k-1} \geq \phi_{k-1}$ & $\ket{1}\leftrightarrow \ket{k-1}$ & - & - & - & \dots & - & - \\
			$\psi_{k} \geq \phi_{k}$ & $\ket{1}\leftrightarrow \ket{k}$ & - & - & - & \dots & - & - \\
			$\psi_d \geq \phi_d$ & $\ket{1}\leftrightarrow \ket{d}$ & $\ket{k+1}\leftrightarrow \ket{d}$ & $\ket{k+2}\leftrightarrow \ket{d}$ & $\ket{k+3}\leftrightarrow \ket{d}$ & \dots & $\ket{d-2}\leftrightarrow \ket{d}$ & $\ket{d-1}\leftrightarrow \ket{d}$ \\
		\end{tabular}
	\end{ruledtabular}
\end{table*}

Note that we applied protocol I, to check and verify its validity, to all possible $d'$-level  ($d'=2,3,4,5,6,7$) sources and target coherent states (and also some special cases of higher level systems) and obtained the explicit solutions, i.e.,
sets of permutations, probabilities, and Kraus operators, in line with the above discussions. Hence, we conclude that the protocol we have presented solves the problem of single-step deterministic transformations of coherent states via incoherent operations.


\subsubsection{Examples of generalized solutions using of protocol I}

\begin{figure}
	\centering
	\includegraphics[width=0.35\textwidth]{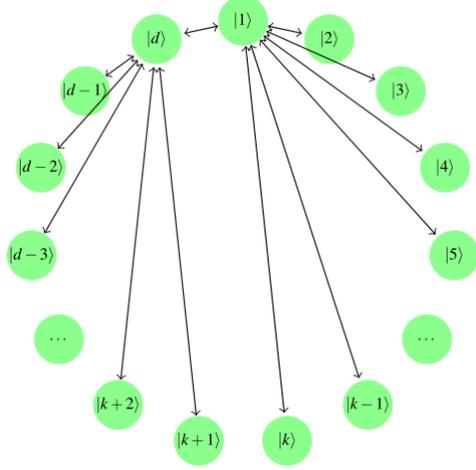}
	\caption{Pictorial representation of permutations for the case $\psi_1 \leq \phi_1$, $\psi_2 \geq \phi_2$,  $\psi_3 \geq \phi_3$, $\dots$, $\psi_{k-1} \geq \phi_{k-1}$, $\psi_{k} \geq \phi_{k}$, $\psi_{k+1} \leq \phi_{k+1}$, $\psi_{k+2} \leq \phi_{k+2}$, $\dots$, $\psi_{d-2} \leq \phi_{d-2}$, $\psi_{d-1} \leq \phi_{d-1}$, $\psi_d \geq \phi_d$. of $d$-level systems.}
\label{gen-fig3}
\end{figure}

Although finding the SPs (and hence finding probabilities) is easy, constructing a general form of probabilities is a highly non-trivial problem, and the problem becomes exponentially difficult as the dimension gets greater. However, we are able to extrapolate a complete solution for some special cases of $d$-level systems. As an example, we consider the case $\psi_1 \leq \phi_1$, $\psi_2 \geq \phi_2$,  $\psi_3 \geq \phi_3$, $\dots$, $\psi_{k-1} \geq \phi_{k-1}$, $\psi_{k} \geq \phi_{k}$, $\psi_{k+1} \leq \phi_{k+1}$, $\psi_{k+2} \leq \phi_{k+2}$, $\dots$, $\psi_{d-2} \leq \phi_{d-2}$, $\psi_{d-1} \leq \phi_{d-1}$, $\psi_d \geq \phi_d$ (see Fig. \eqref{gen-fig3})and construct the table for the  possible permutations as listed in Table \ref{gen}.
The Kraus operators are given by
\begin{eqnarray}
K^i=U^i\Big(\sqrt{p^i}\sum_{j=1}^{d}\frac{c_{ij}}{\psi_j}\ket{j}\bra{j}\Big) \quad (i=1,\dots,d), \
\end{eqnarray}
where $\sum_{i=1}^{d} {K^i}^{\dag}K^i=I_d$ and $p^i=Tr(K^i \rho_{\psi} {K^i}^{\dag})$.
A single SP, for the complete solution, is sufficient for this case where it is found to be
\begin{widetext}
\begin{eqnarray}\begin{aligned}\label{case3n}
U=\Big\{&I_d, \ket{k+1}\leftrightarrow \ket{d}, \ket{k+2}\leftrightarrow \ket{d}, \ket{k+3}\leftrightarrow \ket{d}, \dots, \ket{d-2}\leftrightarrow \ket{d}, \ket{d-1}\leftrightarrow \ket{d}, \ket{1}\leftrightarrow \ket{2},  \ket{1}\leftrightarrow \ket{3}, \\ & \ket{1}\leftrightarrow \ket{4}, \dots, \ket{1}\leftrightarrow \ket{k-2}, \ket{1}\leftrightarrow \ket{k-1},  \ket{1}\leftrightarrow \ket{k},  \ket{1}\leftrightarrow \ket{d} \Big\}.
\end{aligned}\end{eqnarray}
The probabilities are found to be
\begin{eqnarray}\begin{aligned}
\{p^i\}^{i=1,2,\dots,d}=\Big\{
p^1&=1-\sum_{i=2}^{d} p^i, \quad
p^2
=\frac{\alpha_{k+1}}{\gamma_{(k+1)d}}, \quad
p^3
=\frac{\alpha_{k+2}}{\gamma_{(k+2)d}}, \quad
p^4
=\frac{\alpha_{k+3}}{\gamma_{(k+3)d}}, \quad \dots, \quad
p^{d-k-1}
=\frac{\alpha_{d-2}}{\gamma_{(d-2)d}}, \\
p^{d-k}
&=\frac{\alpha_{d-1}}{\gamma_{(d-1)d}}, \quad
p^{d-k+1}
=\frac{\beta_{2}}{\gamma_{12}}, \quad
p^{d-k+2}
=\frac{\beta_{3}}{\gamma_{13}}, \quad \dots, \quad
p^{d-2}
=\frac{\beta_{k-1}}{\gamma_{1(k-1)}}, \quad
p^{d-1}
=\frac{\beta_{k}}{\gamma_{1k}}, \quad
p^d
=\frac{\alpha_{12\dots k}}{\gamma_{1d}}\Big\},
\end{aligned}\end{eqnarray}
\end{widetext}
where $\sum_{j=1}^{k}\phi_{j}^2 \geq \sum_{j=1}^{k}\psi_{j}^2$ (or equivalently $\alpha_{12\dots k} \geq 0$) follows from majorization. The resulting table consists of a single row for the case $\psi_1 \leq \phi_1$, $\psi_2 \leq \phi_2$,  $\psi_3 \leq \phi_3$, $\dots$, $\psi_{d-1} \leq \phi_{d-1}$, $\psi_{d} \geq \phi_{d}$ . The set of permutations and the probabilities in that case turn out to be
\begin{eqnarray}\begin{aligned}
U=\big\{&I_d, \ket{1}\leftrightarrow \ket{d}, \ket{2}\leftrightarrow \ket{d}, \ket{3}\leftrightarrow \ket{d}, \dots, \\ &\ket{d-3}\leftrightarrow \ket{d}, \ket{d-2}\leftrightarrow \ket{d}, \ket{d-1}\leftrightarrow \ket{d}\},
\end{aligned}
\end{eqnarray}
\begin{eqnarray}
p^{1}=1-\sum_{i=2}^{d}p^i, \quad \{p^i\}^{i=2,\dots,d}=\frac{\alpha_{i-1}}{\gamma_{(i-1)d}}=\frac{\phi_{i-1}^2 - \psi_{i-1}^2}{\phi_{i-1}^2 - \phi_{d}^2}, \quad
\end{eqnarray}
respectively. Similarly, the resulting table consists of a single column for the case $\psi_1 \leq \phi_1$, $\psi_2 \geq \phi_2$,  $\psi_3 \geq \phi_3$, $\dots$, $\psi_{d-1} \geq \phi_{d-1}$, $\psi_{d} \geq \phi_{d}$.  The set of permutations and the probabilities in that case turn out to be
\begin{eqnarray}\begin{aligned}
U=\big\{&I_d, \ket{1}\leftrightarrow \ket{2}, \ket{1}\leftrightarrow \ket{3}, \ket{1}\leftrightarrow \ket{4}, \dots, \\
&\ket{1}\leftrightarrow \ket{d-2}, \ket{1}\leftrightarrow \ket{d-1}, \ket{1}\leftrightarrow \ket{d} \big\},
\end{aligned}
\end{eqnarray}
\begin{eqnarray}
p^{1}=1-\sum_{i=2}^{d}p^i, \quad \{p^i\}^{i=2,\dots,d}=\frac{\beta_{i}}{\gamma_{1i}}=\frac{\psi_{i}^2 - \phi_{i}^2}{\phi_1^2-\phi_{i}^2}, \quad
\end{eqnarray}
respectively. Besides these, there are also different cases which are generalizable. If a table of permutations, for instance, has $d-1$ permutations then for such cases probabilities can easily be found while a single SP is sufficient. Also, the number of a single SP is $d-1$ for $d$-level systems where the other $2^{d-2}-d+1$ ($d>3$) cases split into subcases, and therefore, the complete solutions for these cases require more than one SP. By using protocol-I, one can obtain all SPs of a complete set.

\subsection{Protocol II}

We now present another protocol for coherence transformations similar to the one for the LOCC deterministic transformations of bipartite entangled pure states \cite{deterministic}. We use $d'$-dimensional subspace solutions ($d'<d$), obtained by protocol I, to achieve the transformation $\ket{\psi}\bra{\psi} \rightarrow \ket{\phi}\bra{\phi}$ as a sequence of transformations of coherent states, that is, $\ket{\psi}\bra{\psi}$ $\rightarrow$ $\ket{\eta_1}\bra{\eta_1}$ $\rightarrow \dots \rightarrow$ $\ket{\eta_k}\bra{\eta_k}$ $\rightarrow$ $\ket{\phi}\bra{\phi}$, where $\ket{\eta_i}$ are intermediate coherent states satisfying the majorization conditions $\mu{({\psi})}$ $\prec$ $\mu{({\eta_1})}$ $\prec$ $\dots$ $\prec$ $\mu{({\eta_k})}$ $\prec$ $\mu{({\phi})}$. We stress that the most crucial point of the sequence of coherent states transformations is to preserve the majorization condition for the entire transformation. In each step of $\ket{\psi}\bra{\psi} \rightarrow \ket{\phi}\bra{\phi}$ we aim to obtain some coefficients of the final coherent state $\ket{\phi}$ using $d'$-level subspace solutions (we move closer to the final coherent state step-by-step).

In the following protocol, we use five-level ($d'=5$) subspace solutions and transform four coefficients of the existing coherent state to the coefficients of the final coherent state starting from smaller ones. This protocol is implementable for transforming any coefficients satisfying the majorization condition given by the inequalities $\sum_{j=1}^{k}{\psi_j}^{2}\leq\sum_{j=1}^{k}{\phi_j}^{2}$, and an equivalent form of the majorization condition is given by
\begin{eqnarray}
\sum_{j=k+1}^{d}\psi_j^2 \geq \sum_{j=k+1}^{d}\phi_j^2,
\end{eqnarray}
where $k=1,2,\dots,d-1$ and $\sum_{j=1}^{d}\psi_j^2 = \sum_{j=1}^{d}\phi_j^2=1$. We have coherent states $\ket{\psi}$, $\ket{\eta_1}$, $\ket{\eta_2}$, \dots, $\ket{\eta_k}$ and $\ket{\phi}$ such that
\begin{eqnarray}\begin{aligned}
\mu{(\psi)}&=\big(\psi_1^2, \dots, \psi_{d-5}^2, \psi_{d-4}^2, \psi_{d-3}^2, \psi_{d-2}^2, \psi_{d-1}^2, \psi_d^2\big)^T, \\
\mu{({\eta_1})}&=\big(\psi_1^2, \dots, \psi_{d-6}^2, \psi_{d-5}^2, \phi'^2_{d-4}, \phi_{d-3}^2, \phi_{d-2}^2, \phi_{d-1}^2, \phi_d^2\big)^T, \\
\mu{({\eta_2})}&=\big(\psi_1^2, \dots, \psi_{d-9}^2, \phi'^2_{d-8}, \phi_{d-7}^2, \phi_{d-6}^2, \dots, \phi_{d-1}^2, \phi_d^2\big)^T, \\
& \vdots  \\
\mu{({\eta_k})}&=\big(\psi^2_1, \dots, \psi^2_{m-1}, \psi^2_m, \phi'^2_{m+1}, \phi^2_{m+2}, \dots, \phi_{d-1}^2, \phi_d^2\big)^T, \\
\mu{({\phi})}&=\big(\phi_1^2, \phi_2^2, \phi_3^2, \dots, \phi_{d-3}^2, \phi_{d-2}^2, \phi_{d-1}^2, \phi_d^2\big)^T,
\end{aligned}\end{eqnarray}
where $1\leq m \leq 4$. Here, for the first step of the entire transformation, we have $\sum_{j=d-4}^{d}\psi_{j}^2$ $=$ $\sum_{j=d-3}^{d}\phi_{j}^2$ $+$ $\phi^{'2}_{d-4}$, and the majorization condition implies those  $\sum_{j=d-3}^{d}\psi_{j}^2$ $\geq$ $\sum_{j=d-3}^{d}\phi_{j}^2$ and $\psi_{d-4}^2$ $\leq$ $\phi^{'2}_{d-4}$. Additionally, if one starts to transform smaller coefficients, there is a probability of obtaining the relations $\phi'^2_{l} > \phi_{1}^2$ and $\phi'^2_{l} > \psi_{k}^2$ between the coefficients of the coherent states where  $k=1,\dots,(l-1)$ and $l=(d-4), (d-8),\dots,(m+1)$, which may violate the majorization condition, and no further deterministic transformation to the final state would be possible. Therefore, while we construct protocol II, we impose that the intermediate states satisfy $\phi'^2_{l} \leq \psi_{l-1}^2$ in any step of the entire transformation. In the following, however, we are going to discuss the cases by an explicit example where the intermediate states satisfy  $\phi'^2_{l} > \phi_{1}^2$ and $\phi'^2_{l} > \psi_{l-1}^2$ or more generally, $\phi'^2_{l} \geq \psi_{x}^2$ [$x=1,2,\dots,(l-1)$], and we explain how one can overcome this problem. Thus, in each step of the entire transformation the majorization condition is preserved, i.e., $\mu{({\psi})}$ $\prec$ $\mu{({\eta_1})}$ $\prec$ $\dots$ $\prec$ $\mu{({\eta_k})}$ $\prec$ $\mu{({\phi})}$, under the assumption $\phi'^2_{l} \leq \psi_{l-1}^2$.

\begin{figure}[!htbp]
	\centering
	\includegraphics[width=0.15\textwidth]{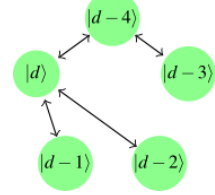}
	\caption{Pictorial representation of the permutations given in Table \ref{table-d'5}. Here, four-element combination of the permutations together with $I_5$ (a single SP)  is sufficient to constitute a complete set.}
	\label{figure-d'5}
\end{figure}

This protocol consists of deterministic transformations of $d'$-dimensional (consider $d'=5$) subspace of the initial state \eqref{psigenform} in each step (and for the last step $2\leq d' \leq 5$). We give the explicit solutions for the first two transformations, $\ket{\psi}\bra{\psi}$ $\rightarrow$ $\ket{\eta_1}\bra{\eta_1}$ and $\ket{\eta_1}\bra{\eta_1}$ $\rightarrow$ $\ket{\eta_2}\bra{\eta_2}$, and the last transformation, $\ket{\eta_k}\bra{\eta_k}$ $\rightarrow$ $\ket{\phi}\bra{\phi}$, for illustrative purposes, and other steps of complete transformation can be treated similarly. The initial coherent state $\ket{\psi}$ given in Eq. \eqref{psigenform} can be written in the form
\begin{eqnarray}\begin{aligned}
\ket{\psi}=\sum_{j=1}^{d-5} \psi_{j}\ket{j}+\eta_1 \ket{\chi},
\end{aligned}\end{eqnarray}
where $\ket{\chi}=\sum_{i=d-4}^{d} \chi_{i}\ket{i}$ is a normalized coherent state  ($\sum_{i=d-4}^{d} {\chi_{i}}^2=1$). Here, we have
$\eta_1\chi_d=\psi_d$, $\eta_1\chi_{d-1}=\psi_{d-1}$, $\eta_1\chi_{d-2}=\psi_{d-2}$, $\eta_1\chi_{d-3}=\psi_{d-3}$, and $\eta_1\chi_{d-4}=\psi_{d-4}$. The intermediate coherent states are given by
\begin{equation}\label{fk}
\ket{\eta_{l}}=\sum_{j=1}^{d-4l-1} \psi_j \ket{j}+\phi'_{d-4l}\ket{d-4l}+\sum_{j=d-4l+1}^{d} \phi_j \ket{j},
\end{equation}
where
\begin{equation}\label{n1}
\phi'_{d-4l}=\sqrt{\psi_{d-4l}^2+\sum_{j=d-4l+1}^{d} \left(\psi_j^2-\phi_j^2\right)},
\end{equation}
with $1\leq l \leq\lfloor {(d-2)}/{4} \rfloor$, and we assume $\phi'_{d-4l}\leq\psi_{d-4l-1}$. In each step, four more coefficients of the target state $\ket{\phi}$ are obtained, i.e., the smallest $4l$ coefficients of $\ket{\eta_{l}}$ and $\ket{\phi}$ are equal. For the first transformation $\ket{\psi}\bra{\psi}$ $\rightarrow$ $\ket{\eta_1}\bra{\eta_1}$, the pure coherent state $\ket{\eta_1}$ is of the form
\begin{eqnarray}\begin{aligned}
\ket{\eta_1}=\sum_{j=1}^{d-5} \psi_{j}\ket{j}+\eta_1 \ket{\tilde{\chi}},
\end{aligned}\end{eqnarray}
where $\ket{\tilde{\chi}}=\sum_{i=d-4}^{d} \tilde{\chi}_{i}\ket{i}$ is a normalized coherent state and
$\sum_{j=1}^{d-5} {\psi_{j}}^2+{\eta_1}^2=1$. We have
$\eta_1\tilde{\chi}_d=\phi_d$, $\eta_1\tilde{\chi}_{d-1}=\phi_{d-1}$, $\eta_1\tilde{\chi}_{d-2}=\phi_{d-2}$, $\eta_1\tilde{\chi}_{d-3}=\phi_{d-3}$, and $\eta_1\tilde{\chi}_{d-4}=\phi'_{d-4}$, and the assumption is $\phi'_{d-4}\leq\psi_{d-5}$ [therefore $\mu{(\chi)}$ $\prec$ $\mu{(\tilde{\chi})}$].
The Kraus operators are given by
\begin{eqnarray}
K_{i1}=\sum_{j=1}^{d-5}\sqrt{p_i}\ket{j}\bra{j}\oplus
\tilde{K}_{i1},
\end{eqnarray}
where $\sum_{i=1}^{5}K_{i1}^{\dag}K_{i1}=I_d$ and $\sum_{i=1}^{5}\tilde{K}_{i1}^{\dag}\tilde{K}_{i1}=I_5$. Here, the Kraus operators $\tilde{K}_{i1}$ are written as
\begin{eqnarray}
\tilde{K}_{i1}=U_{i1}\Big(\sqrt{p_i}\sum_{j=d-4}^{d}\frac{z_{ij}}{\chi_j}\ket{j}\bra{j}\Big),
\end{eqnarray}
where $\tilde{K}_{i1}\ket{\chi}=\sqrt{p_i}\ket{\tilde{\chi}}$. Both $\ket{\chi}$ and $\ket{\tilde{\chi}}$ are five-level states, so we use solutions for $d'=5$. There are eight possible cases for coefficients of coherent states except the lowest and greatest ones, and the set of Kraus operators $\{\tilde{K}_{i1}\}_{i=1,\dots,5}$ is chosen accordingly.
\begin{table}[!htbp]
	\caption{All possible permutations for the case $\chi_{d-3} \geq \tilde{\chi}_{d-3}$, $\chi_{d-2} \leq \tilde{\chi}_{d-2}$ and $\chi_{d-1} \leq \tilde{\chi}_{d-1}$ of five-level coherent states   $\ket{\chi}$ and $\ket{\tilde{\chi}}$, where $\chi_{d-4} \leq \tilde{\chi}_{d-4}$ and $\chi_d \geq \tilde{\chi}_d$ follow from the majorization condition.}
	\label{table-d'5}
	\centering
	\begin{ruledtabular}
		\begin{tabular}{c c c c}
			& $\chi_{d-4} \leq \tilde{\chi}_{d-4}$ & $\chi_{d-2} \leq \tilde{\chi}_{d-2}$ & $\chi_{d-1} \leq \tilde{\chi}_{d-1}$   \\ [0.5ex] \hline
			$\chi_{d-3} \geq \tilde{\chi}_{d-3}$ & $\ket{d-4}\leftrightarrow \ket{d-3}$ & - & -  \\
			$\chi_d \geq \tilde{\chi}_d$ & $\ket{d-4}\leftrightarrow \ket{d}$ & $\ket{d-2}\leftrightarrow \ket{d}$ &  $\ket{d-1}\leftrightarrow \ket{d}$  \\
		\end{tabular}
	\end{ruledtabular}
\end{table}
For instance, let us consider the case $\chi_{d-3} \geq \tilde{\chi}_{d-3}$, $\chi_{d-2} \leq \tilde{\chi}_{d-2}$, and $\chi_{d-1} \leq \tilde{\chi}_{d-1}$  where $\chi_{d-4} \leq \tilde{\chi}_{d-4}$ and $\chi_d \geq \tilde{\chi}_d$ follow from the majorization condition. In Table \ref{table-d'5}, there are four permutations, and these constitute the SP together with identity transformation $I_5$, i.e., a single set is sufficient for the complete solution. Figure \ref{figure-d'5} provides the pictorial representation of the permutations given in Table \ref{table-d'5}. The set of permutations is obtained as  $\{U_{i1}\}_{i=1,\dots,5}=\big\{I_5, \ket{d-4}\leftrightarrow \ket{d}, \ket{d-1}\leftrightarrow \ket{d}, \ket{d-2}\leftrightarrow \ket{d}, \ket{d-4}\leftrightarrow \ket{d-3} \big\}$, where $I_5=\sum_{u=d-4}^{d}\ket{u}\bra{u}$.
Then, if we perform Kraus operators, we obtain
\begin{widetext}
\begin{eqnarray}\begin{aligned}
K_{i1}\ket{\psi}&=\bigg(\sum_{j=1}^{d-5}\sqrt{p_i}\ket{j}\bra{j}\oplus U_{i1}\Big(\sqrt{p_i}\sum_{j=d-4}^{d}\frac{z_{ij}}{\chi_j}\ket{j}\bra{j}\Big)\bigg)\bigg(\sum_{k=1}^{d-5} \psi_{k}\ket{k}+\eta_1 \sum_{k=d-4}^{d} \chi_{k}\ket{k}\bigg)\\
&=\sqrt{p_i}\Big(\sum_{j=1}^{d-5}\sum_{k=1}^{d-5}\psi_{k}\ket{j}\braket{j \mid k}+
\sum_{j=1}^{d-5}\sum_{k=d-4}^{d}\psi_{k}\ket{j}\braket{j \mid k}+
U_{i1}\sum_{j=d-4}^{d}\sum_{k=1}^{d-5}\frac{z_{ij}}{\chi_j}\psi_{k}\ket{j}\braket{j\mid k}+ U_{i1}\eta_1\sum_{j=d-4}^{d}\sum_{k=d-4}^{d}\frac{z_{ij}}{\chi_j}\chi_{k}\ket{j}\braket{j\mid k}\Big) \\
&=\sqrt{p_i}\Big(\sum_{j=1}^{d-5}\sum_{k=1}^{d-5}\psi_{k}\ket{j}\braket{j \mid k}+ U_{i1}\eta_1\sum_{j=d-4}^{d}\sum_{k=d-4}^{d}\frac{z_{ij}}{\chi_j}\chi_{k}\ket{j}\braket{j\mid k}\Big) \\
&=\sqrt{p_i}\Big(\sum_{j=1}^{d-5}\psi_{j}\ket{j}+ \eta_1 U_{i1} \sum_{j=d-4}^{d}z_{ij}\ket{j}\Big) \\
&=\sqrt{p_i} \ket{\eta_1}.
\end{aligned}\end{eqnarray}
\end{widetext}
Then, from $K_{i1} \ket{\psi}=\sqrt{p_i}\ket{\eta_1}$ it follows
\begin{eqnarray}
\begin{aligned}\sum_{i=1}^{5} K_{i1}\ket{\psi}\bra{\psi}K_{i1}^{\dag}
=\sum_{i=1}^{5}p_i\ket{\eta_1}\bra{\eta_1}
=\ket{\eta_1}\bra{\eta_1}.
\end{aligned}
\end{eqnarray}
The condition $\sum_{i=1}^{5}\tilde{K}_{i1}^{\dag}\tilde{K}_{i1}=I_5$ implies the following linear equations
\begin{eqnarray}\begin{aligned}
\sum_{i=1}^{5} {p_i} {z^2_{ij}}=\chi^2_j, \quad (j=(d-4),(d-3),\dots,d),
\end{aligned}
\end{eqnarray}
whose solutions for $p_i$ give the probabilities where
$(z_{i(d-4)}, z_{i(d-3)},\dots, z_{id})^{T}=U_{i1}^{\dag}(\phi'_{d-4}, \phi_{d-3}, \phi_{d-2}, \phi_{d-1}, \phi_{d})^{T}$. We denote $z_{ij}$ as the $(ij)$th element of the matrix $z$ given by
\begin{eqnarray}
z=\left(\begin{array}{ccccc}  \phi'_{d-4} & \phi_{d-3} & \phi_{d-2} & \phi_{d-1} & \phi_{d} \\ \phi_{d} & \phi_{d-3} & \phi_{d-2} & \phi_{d-1} & \phi'_{d-4} \\ \phi'_{d-4} & \phi_{d-3} & \phi_{d-2} & \phi_{d} & \phi_{d-1} \\ \phi'_{d-4} & \phi_{d-3} & \phi_{d} & \phi_{d-1} & \phi_{d-2} \\ \phi_{d-3} & \phi'_{d-4} & \phi_{d-2} & \phi_{d-1} & \phi_{d}  \end{array}\right).
\end{eqnarray}
The probabilities are found to be
\begin{eqnarray}\begin{aligned}
p_1&=1-\sum_{i=1}^4 p_i, \quad
p_2=\frac{\tilde{\chi}^2_{d-4}+\tilde{\chi}_{d-3}^2-\chi^2_{d-4}
-\chi_{d-3}^2}{\tilde{\chi}^2_{d-4}-\tilde{\chi}_{d}^2}, \\ p_3&=\frac{\tilde{\chi}_{d-1}^2-\chi_{d-1}^2}{\tilde{\chi}_{d-1}^2-\tilde{\chi}_d^2}, \quad
p_4=\frac{\tilde{\chi}_{d-2}^2-\chi_{d-2}^2}{\tilde{\chi}_{d-2}^2-\tilde{\chi}_d^2}, \\
p_5&=\frac{\chi_{d-3}^2-\tilde{\chi}_{d-3}^2}{\tilde{\chi}^2_{d-4}-\tilde{\chi}_{d-3}^2},
\end{aligned}\end{eqnarray}
where $\tilde{\chi}^2_{d-4}+\tilde{\chi}_{d-3}^2\geq\chi^2_{d-4}
+\chi_{d-3}^2$ due to the majorization condition. Thus, we obtain the state $\ket{\eta_1}\bra{\eta_1}$. Now, in the next step, we obtain the transformation $\ket{\eta_1}\bra{\eta_1}$ $\rightarrow$ $\ket{\eta_2}\bra{\eta_2}$. The pure coherent state $\ket{\eta_1}$ is written in the form
\begin{eqnarray}
\ket{\eta_1}&=&\sum_{j=1}^{d-5}\psi_{j}\ket{j}+\phi'_{d-4}\ket{d-4}+\sum_{j=d-3}^{d}\phi_{j}\ket{j}
\nonumber \\
&=&\sum_{j=1}^{d-9} \psi_{j}\ket{j}+\eta_2\ket{\varphi}+\sum_{j=d-3}^{d}\phi_{j}\ket{j},
\end{eqnarray}
where $\ket{\varphi}=\sum_{j=d-8}^{d-4}{\varphi_{j}}\ket{j}$ is a normalized coherent state ($\sum_{i=d-8}^{d-4} {\varphi_{i}}^2=1$), $\eta_2\varphi_{d-8}=\psi_{d-8}$, $\eta_2\varphi_{d-7}=\psi_{d-7}$,
$\eta_2\varphi_{d-6}=\psi_{d-6}$, $\eta_2\varphi_{d-5}=\psi_{d-5}$, and $\eta_2\varphi_{d-4}=\phi'_{d-4}$.
The second intermediate coherent state is given by
\begin{eqnarray}
\ket{\eta_2}=\sum_{j=1}^{d-9} \psi_{j}\ket{j}+\eta_2\ket{\tilde{\varphi}}+\sum_{j=d-3}^{d}\phi_{j}\ket{j},
\end{eqnarray}
where $\ket{\tilde{\varphi}}=\sum_{j=d-8}^{d-4}{\tilde{\varphi}_{j}}\ket{j}$ is a normalized coherent state ($\sum_{i=d-8}^{d-4} {\tilde{\varphi}_{i}}^2=1$), $\eta_2\tilde{\varphi}_{d-7}=\phi_{d-7}$,  $\eta_2\tilde{\varphi}_{d-6}=\phi_{d-6}$, $\eta_2\tilde{\varphi}_{d-5}=\phi_{d-5}$,  $\eta_2\tilde{\varphi}_{d-4}=\phi_{d-4}$, and $\eta_2\tilde{\varphi}_{d-8}=\phi'_{d-8}$ (and the assumption is $\phi'_{d-8}\leq\psi_{d-9}$). Thus, we have $\mu{(\varphi)}$ $\prec$ $\mu{(\tilde{\varphi})}$. The Kraus operators are given by
\begin{eqnarray}
K_{i2}=\sum_{j=1}^{d-9}\sqrt{p_i}\ket{j}\bra{j}\oplus \tilde{K}_{i2} \oplus\sum_{j=d-3}^{d}\sqrt{p_i}\ket{j}\bra{j},
\end{eqnarray}
where $\sum_{i=1}^{5}K_{i2}^{\dag}K_{i2}=I_d$ and $\sum_{i=1}^{5}\tilde{K}_{i2}^{\dag}\tilde{K}_{i2}=I_5$. Here, the Kraus operators $\{\tilde{K}_{i2}\}_{i=1,\dots,5}$ are given by
\begin{eqnarray}
\tilde{K}_{i2}=U_{i2}\Big(\sqrt{p_i}\sum_{j=d-8}^{d-4}\frac{z_{ij}'}{\varphi_j}\ket{j}\bra{j}\Big),
\end{eqnarray}
where they are chosen according to the eight possible cases of five-level systems to make the transformation $\tilde{K}_{i2}\ket{\varphi}=\sqrt{p_i}\ket{\tilde{\varphi}}$. For instance, let us consider the case $\varphi_{d-7} \geq \tilde{\varphi}_{d-7}$, $\varphi_{d-6} \geq \tilde{\varphi}_{d-6}$ and $\varphi_{d-5} \geq \tilde{\varphi}_{d-5}$.
\begin{table}[!htbp]
	\caption{All possible permutations for the case $\varphi_{d-7} \geq \tilde{\varphi}_{d-7}$, $\varphi_{d-6} \geq \tilde{\varphi}_{d-6}$, and $\varphi_{d-5} \geq \tilde{\varphi}_{d-5}$  of five-level coherent states   $\ket{\varphi}$ and $\ket{\tilde{\varphi}}$ where $\varphi_{d-8} \leq \tilde{\varphi}_{d-8}$ and $\varphi_{d-4} \geq \tilde{\varphi}_{d-4}$ follow from the majorization condition.}
	\label{table2-d'5}
	\centering
	\begin{ruledtabular}
		\begin{tabular}{c c c c}
			& $\varphi_{d-8} \leq \tilde{\varphi}_{d-8}$  \\ [0.5ex] \hline
			$\varphi_{d-7} \geq \tilde{\varphi}_{d-7}$ & $\ket{d-8}\leftrightarrow \ket{d-7}$   \\
			$\varphi_{d-6} \geq \tilde{\varphi}_{d-6}$ & $\ket{d-8}\leftrightarrow \ket{d-6}$  \\
			$\varphi_{d-5} \geq \tilde{\varphi}_{d-5}$ & $\ket{d-8}\leftrightarrow \ket{d-5}$  \\
			$\varphi_{d-4} \geq \tilde{\varphi}_{d-4}$ & $\ket{d-8}\leftrightarrow \ket{d-4}$ \\
		\end{tabular}
	\end{ruledtabular}
\end{table}
In Table \ref{table2-d'5}, there are four permutations, and these, together with identity transformation $I_5$, constitute the set of permutations (a single set is sufficient).
\begin{figure}[!htbp]
	\centering
	\includegraphics[width=0.15\textwidth]{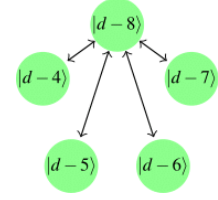}
	\caption{Pictorial representations of the four-element combination of the permutations given in Table \ref{table2-d'5}.}
	\label{figure2-d'5}
\end{figure}
Figure \ref{figure2-d'5} provides the pictorial representation of the permutations given in Table \ref{table2-d'5}. The SP is obtained as  $\{U_{i1}\}_{i=1,\dots,5}=\big\{I_5, \ket{d-8}\leftrightarrow \ket{d-7}, \ket{d-8}\leftrightarrow \ket{d-6}, \ket{d-8}\leftrightarrow \ket{d-5}, \ket{d-8}\leftrightarrow \ket{d-4} \big\}$ where $I_5=\sum_{u=d-8}^{d-4}\ket{u}\bra{u}$.
Then, from $K_{i2} \ket{\eta_1}=\sqrt{p_i}\ket{\eta_2}$ it follows
\begin{eqnarray}\begin{aligned}
\sum_{i=1}^{5}K_{i2} \ket{\eta_1}\bra{\eta_1}K_{i2}^{\dag}
=\sum_{i=1}^{5}p_i\ket{\eta_2}\bra{\eta_2}
=\ket{\eta_2}\bra{\eta_2}.
\end{aligned}\end{eqnarray}
The condition $\sum_{i=1}^{5}\tilde{K}_{i2}^{\dag}\tilde{K}_{i2}=I_5$ implies the following linear equations
\begin{eqnarray}\begin{aligned}
\sum_{i=1}^{5} {p_i} {z'^2_{ij}}=\varphi^2_j, \quad (j=(d-8),(d-7),\dots,(d-4)),
\end{aligned}
\end{eqnarray}
whose solutions for $p_i$ give the probabilities. We denote $z'_{ij}$ as the $(ij)$th element of the matrix $z'$ given by
\begin{eqnarray}
z'=\left(\begin{array}{ccccc}  \phi'_{d-8} & \phi_{d-7} & \phi_{d-6} & \phi_{d-5} & \phi_{d-4} \\ \phi_{d-7}  & \phi'_{d-8} & \phi_{d-6} & \phi_{d-5} & \phi_{d-4} \\ \phi_{d-6} & \phi_{d-7} & \phi'_{d-8} & \phi_{d-5} & \phi_{d-4} \\ \phi_{d-5} & \phi_{d-7} & \phi_{d-6} & \phi'_{d-8} & \phi_{d-4} \\ \phi_{d-4} & \phi_{d-7} & \phi_{d-6} & \phi_{d-5} & \phi'_{d-8}  \end{array}\right).
\end{eqnarray}
The probabilities are found to be
\begin{eqnarray}\begin{aligned}
p_1&=1-\sum_{i=1}^4 q_i, \
p_2=\frac{\varphi_{d-7}^2-\tilde{\varphi}^2_{d-7}}{\tilde{\varphi}^2_{d-8}-\tilde{\varphi}_{d-7}^2}, \
p_3=\frac{\varphi_{d-6}^2-\tilde{\varphi}^2_{d-6}}{\tilde{\varphi}_{d-8}^2-\tilde{\varphi}_{d-6}^2}, \\
p_4&=\frac{\varphi_{d-5}^2-\tilde{\varphi}^2_{d-5}}{\tilde{\varphi}^2_{d-8}-\tilde{\varphi}_{d-5}^2}, \
p_5=\frac{\varphi_{d-4}^2-\tilde{\varphi}^2_{d-4}}{\tilde{\varphi}_{d-8}^2-\tilde{\varphi}_{d-4}^2},
\end{aligned}\end{eqnarray}
Thus, we obtained the state $\ket{\eta_2}\bra{\eta_2}$ where
\begin{eqnarray}
\ket{\eta_2}=\sum_{j=1}^{d-9} \psi_{j}\ket{j}+\phi'_{d-8}\ket{d-8}+\sum_{j=d-7}^{d}\phi_{j}\ket{j}.
\end{eqnarray}
For the next to the last transformation of the entire transformation we have the coherent states given by
\begin{eqnarray}
\ket{\eta_{k-1}}&=&\sum_{j=1}^{m} \psi_{j}\ket{j}+\eta_{k}\sum_{j=m+1}^{m+5}{\omega_{j}}\ket{j}+\sum_{j=m+6}^{d}\phi_{j}\ket{j},
\end{eqnarray}
and
\begin{eqnarray}
\ket{\eta_{k}}&=&\sum_{j=1}^{m} \psi_{j}\ket{j}+\eta_{k}\sum_{j=m+1}^{m+5}{\tilde{\omega}_{j}}\ket{j}+\sum_{j=m+6}^{d}\phi_{j}\ket{j},
\end{eqnarray}
where $1\leq m \leq 4$, $\eta_k\omega_{m+1}=\psi_{m+1}$, $\eta_k\omega_{m+2}=\psi_{m+2}$,
$\eta_k\omega_{m+3}=\psi_{m+3}$, $\eta_k\omega_{m+4}=\psi_{m+4}$, and $\eta_k\omega_{m+5}=\phi'_{m+5}$.
Since our aim is the obtain four more coefficients of the target state $\ket{\phi}$, the coefficients are chosen to be
$\eta_k\tilde{\omega}_{m+2}=\phi_{m+2}$, $\eta_k\tilde{\omega}_{m+3}=\phi_{m+3}$,
$\eta_k\tilde{\omega}_{m+4}=\phi_{m+4}$, $\eta_k\tilde{\omega}_{m+5}=\phi_{m+5}$, and $\eta_k\tilde{\omega}_{m+1}=\phi'_{m+1}$ (and the assumption is $\phi'_{m+1}\leq\psi_{m}$). Thus, we obtain the intermediate state transformations, i.e., $\ket{\eta_{k-1}}\bra{\eta_{k-1}} \rightarrow \ket{\eta_{k}}\bra{\eta_{k}}$. The last transformation of the entire transformation is $\ket{\eta_{k}}\bra{\eta_{k}} \rightarrow \ket{\phi}\bra{\phi}$. The state $\ket{\eta_{k}}$ is written as
\begin{eqnarray}
\ket{\eta_{k}}=\sum_{j=1}^{m} \psi_{j}\ket{j}+\phi'_{m+1}\ket{m+1}+\sum_{j=m+2}^{d}\phi_{j}\ket{j},
\end{eqnarray}
where $1\leq m \leq 4$, and the final state is $\ket{\phi}=\sum_{j=1}^{d}\phi_{j}\ket{j}$. The last transformation is effectively $(m+1)$-dimensional transformation. Let us consider, for instance, the case $m=2$. Then we have
\begin{eqnarray}\begin{aligned}
\ket{\eta_{k}}&=\psi_{1}\ket{1}+\psi_{2}\ket{2}+\phi'_{3}\ket{3}+\sum_{j=4}^{d}\phi_{j}\ket{j},
\end{aligned}\end{eqnarray}
and the Kraus operators are given by
\begin{eqnarray}
K_{i(k+1)}&=&\tilde{K}_{i(k+1)}\oplus\sum_{j=4}^{d}\sqrt{p_i}\ket{j}\bra{j} \nonumber \\
&=&U_{i(k+1)}\sqrt{p_i}\Big(\frac{z_{i1}}{\psi_1}\ket{1}\bra{1}+\frac{z_{i2}}{\psi_2}\ket{2}\bra{2}
+\frac{z_{i3}}{\phi'_3}\ket{3}\bra{3}\Big) \nonumber \\
&&\oplus\sum_{j=4}^{d}\sqrt{p_i}\ket{j}\bra{j},
\end{eqnarray}
where $\sum_{i=1}^{3}{K^{\dag}_{i(k+1)}}{K_{i(k+1)}}=I_d$ and $\sum_{i=1}^{3}{\tilde{K}^{\dag}_{i(k+1)}}{\tilde{K}_{i(k+1)}}=I_3$. Here, we have $\phi'_3 \geq\phi_3$, and one of the inequality relations $\psi_2 \leq(\geq)\phi_2$  can be possible, and the set of Kraus operators $\{\tilde{K}_{i(k+1)}\}$ is chosen accordingly. For instance, for the case $\psi_2 \leq \phi_2$, the set of unitary permutations is obtained as  $\{U_{i(k+1)}\}_{i=1,2,3}=\big\{I_3, \ket{2}\leftrightarrow \ket{3}, \ket{1}\leftrightarrow \ket{3} \big\}$. The condition $\sum_{i=1}^{3}\tilde{K}_{i(k+1)}^{\dag}\tilde{K}_{i(k+1)}=I_3$ implies the following linear equations
\begin{eqnarray}\begin{aligned}
\sum_{i=1}^{3} {p_i} {z^2_{ij}}=\psi^2_j, \quad (j=1,2), \quad \sum_{i=1}^{3} {p_i} {z^2_{i3}}=\phi'^2_3,
\end{aligned}
\end{eqnarray}
whose solutions for $p_i$ give the probabilities. We denote $z_{ij}$ as the $(ij)$th element of the matrix $z$ given by
\begin{eqnarray}
z=\left(\begin{array}{ccc}  \phi_{1} & \phi_{2} & \phi_{3} \\ \phi_{1}  & \phi_{3} & \phi_{2} \\ \phi_{3} & \phi_{2} & \phi_{1}    \end{array}\right).
\end{eqnarray}
The probabilities are found to be $p_1=1-p_2-p_3$, $p_2={(\phi^2_{2}-\psi^2_{2})}/{(\phi^2_{2}-\phi^2_{3})}$ and $p_3={(\phi^2_{1}-\psi^2_{1})}/{(\phi^2_{1}-\phi^2_{3})}$. Then, from $K_{i(k+1)} \ket{\eta_k}=\sqrt{p_i}\ket{\phi}$ it follows
\begin{eqnarray}\begin{aligned}
\sum_{i=1}^{3} K_{i(k+1)}\ket{\eta_k}\bra{\eta_k}K_{i(k+1)}^{\dag}=\sum_{i=1}^{3}p_i\ket{\phi}\bra{\phi}=\ket{\phi}\bra{\phi},
\end{aligned}\end{eqnarray}
where $k+1=\lfloor {(d+2)}/{4} \rfloor$. Thus, we obtain the $d$-level pure coherent states transformations by cascading a sequence of coherence transformations, each corresponding to a single incoherent operation $\Phi[\cdot]$.
The sequence of pure coherent states transformations can be summarized as
\begin{widetext}
\begin{eqnarray}\begin{aligned}
\ket{\psi}\bra{\psi} &\rightarrow \sum_{f=1}^{m+1}K_{f(k+1)}\Bigg(\sum_{r=1}^{5}K_{rk}\bigg(\dots\sum_{l=1}^{5}K_{l3}\Big(\sum_{j=1}^{5}K_{j2}\big(\sum_{i=1}^{5} K_{i1}\ket{\psi}\bra{\psi}K_{i1}^{\dag}\big)K_{j2}^{\dag}\Big)K_{l3}^{\dag}\dots\bigg)K_{rk}^{\dag}\Bigg)K_{f(k+1)}^{\dag} \\
&=\Phi_{(k+1)}\Bigg[\underbrace{\Phi_{(k)}\bigg[\dots\Phi_{(3)}\Big[\underbrace{\Phi_{(2)}
\big[\underbrace{\Phi_{(1)}(\ket{\psi}\bra{\psi})}_{=\ket{\eta_1}\bra{\eta_1}}\big]
}_{=\ket{\eta_2}\bra{\eta_2}}\Big]\dots\bigg]}_{=\ket{\eta_k}\bra{\eta_k}}\Bigg] \\
&=\ket{\phi}\bra{\phi},
\end{aligned}\end{eqnarray}
\end{widetext}
where $1\leq m \leq 4$ and $k+1=\lfloor {(d+2)}/{4} \rfloor$. Consequently, we obtain the entire transformation
$\ket{\psi}\bra{\psi} \rightarrow \ket{\phi}\bra{\phi}$ as $\ket{\psi}\bra{\psi}$ $\rightarrow$ $\ket{\eta_1}\bra{\eta_1}$ $\rightarrow$ $\dots$ $\rightarrow$ $\ket{\eta_k}\bra{\eta_k}$ $\rightarrow$ $\ket{\phi}\bra{\phi}$ in $\lfloor {(d+2)}/{4} \rfloor$ steps via the set of incoherent operations $\{\Phi_{(i)}\}_{i=1,2,\dots,\lfloor {(d+2)}/{4} \rfloor}$. It should be noted that the number of steps can be further reduced by using subspace solutions of $d'$-level where $d'>5$ (actually, number of steps is $\lfloor {(d+d'-3)}/{(d'-1)} \rfloor$).


\subsubsection{Explicit example for protocol II with discussion}

In protocol II we consider deterministic transformations of coherent states by transforming the four smallest nonequal coefficients step-by-step. This procedure requires that the condition $\phi'^2_{l} \leq \psi_{l-1}^2$ is satisfied  in each step of the entire transformation to preserve the majorization condition for intermediate states. However, the condition $\phi'^2_{l} \leq \psi_{l-1}^2$ is not satisfied for some sources and the target states. In all these cases, it is always possible to find intermediate states where lower dimensional transformations can be used. For illustrative purposes, we start with a six-dimensional initial pure coherent state $\ket{\psi}$ and six-dimensional final pure coherent state $\ket{\phi}$ such that
\begin{eqnarray}
\mu{(\psi)}= \frac{1}{53}(11,11, 8, 8, 8, 7)^T,
\end{eqnarray}
\begin{eqnarray}
\mu{(\phi)}=\frac{1}{53}(12, 12, 10, 9, 6, 4)^T,
\end{eqnarray}
respectively. It is obvious that $\mu{(\psi)}\prec\mu{(\phi)}$. For the first step of the entire transformation $\ket{\psi}\bra{\psi}\rightarrow\ket{\phi}\bra{\phi}$, if one starts to transform smaller coefficients, using five-level solutions, then for an intermediate coherent state $\ket{\eta'_1}$ we have
\begin{eqnarray} \begin{aligned}
\mu{(\eta'_1)}= \frac{1}{53}(11, 13, 10, 9, 6, 4)^T,
\end{aligned}\end{eqnarray}
where the last four coefficients of coherent states $\ket{\phi}$ and $\ket{\eta'_1}$ are the same, and also first coefficients of coherent states $\ket{\psi}$ and $\ket{\eta'_1}$ are the same. Since the majorization condition, $\mu{(\eta'_1)}\prec\mu{(\phi)}$, is not satisfied, the deterministic transformation $\ket{\eta'_1}\bra{\eta'_1}\rightarrow\ket{\phi}\bra{\phi}$ is not possible via incoherent operations. This is why we stress that our protocol consists of the intermediate coherent states for which $\phi'^2_{l} \leq \psi_{l-1}^2$ in any step of the entire transformation. On the other hand, for the first step of the complete transformation, if one starts to transform greater coefficients then for an intermediate coherent state $\ket{\eta''_1}$ we have
\begin{eqnarray} \begin{aligned}
\mu{(\eta''_1)}=\frac{1}{53}(12, 12, 10, 9, 3, 7)^T,
\end{aligned}\end{eqnarray}
where the first four coefficients of coherent states $\ket{\phi}$ and $\ket{\eta''_1}$ are the same, and also the last coefficients of coherent states $\ket{\psi}$ and $\ket{\eta''_1}$ are the same.  Since the majorization condition,  $\mu{(\eta''_1)}\prec\mu{(\phi)}$, is not satisfied, the deterministic transformation $\ket{\eta''_1}\bra{\eta''_1}\rightarrow\ket{\phi}\bra{\phi}$ is not possible via incoherent operations. At first glance the transformation $\ket{\psi}\bra{\psi}\rightarrow\ket{\phi}\bra{\phi}$ looks unachievable by protocol II. However, it is always possible to choose intermediate states$-$intermediate states are not unique$-$which do not violate the majorization condition. For the first step of the
complete transformation we consider the pure coherent state $\ket{\eta_1}$ as an intermediate state for which we have
\begin{eqnarray}\begin{aligned}
\mu{(\eta_1)}=\frac{1}{53}(12, 12, 10, 8, 7, 4)^T,
\end{aligned}\end{eqnarray}
where four coefficients (first three and the last) of coherent states $\ket{\phi}$ and $\ket{\eta_1}$ are the same, and also the fourth coefficients of coherent states $\ket{\psi}$ and $\ket{\eta_1}$ are the same. It is obviously seen that the majorization condition is satisfied, $\mu{(\psi)}\prec\mu{(\eta_1)}\prec\mu{(\phi)}$. Then, under the Kraus operators given by
\begin{widetext}
\begin{eqnarray}\begin{aligned}
{K}_{11}&=U_{11}\sqrt{p_1}\Big(\sqrt{\frac{12}{11}}\ket{1}\bra{1}+\sqrt{\frac{12}{11}}\ket{2}\bra{2}
+\sqrt{\frac{5}{4}}\ket{3}\bra{3}+\ket{4}\bra{4}+\sqrt{\frac{7}{8}}\ket{5}\bra{5}+\sqrt{\frac{4}{7}}\ket{6}\bra{6}\Big),\\
{K}_{21}&=U_{21}\sqrt{p_2}\Big(\sqrt{\frac{4}{11}}\ket{1}\bra{1}+\sqrt{\frac{12}{11}}\ket{2}\bra{2}
+\sqrt{\frac{5}{4}}\ket{3}\bra{3}+\ket{4}\bra{4}+\sqrt{\frac{7}{8}}\ket{5}\bra{5}+\sqrt{\frac{12}{7}}\ket{6}\bra{6}\Big), \\
{K}_{31}&=U_{31}\sqrt{p_3}\Big(\sqrt{\frac{12}{11}}\ket{1}\bra{1}+\sqrt{\frac{4}{11}}\ket{2}\bra{2}
+\sqrt{\frac{5}{4}}\ket{3}\bra{3}+\ket{4}\bra{4}+\sqrt{\frac{7}{8}}\ket{5}\bra{5}+\sqrt{\frac{12}{7}}\ket{6}\bra{6}\Big), \\
{K}_{41}&=U_{41}\sqrt{p_4}\Big(\sqrt{\frac{12}{11}}\ket{1}\bra{1}+\sqrt{\frac{12}{11}}\ket{2}\bra{2}
+\sqrt{\frac{7}{8}}\ket{3}\bra{3}+\ket{4}\bra{4}+\sqrt{\frac{5}{4}}\ket{5}\bra{5}+\sqrt{\frac{4}{7}}\ket{6}\bra{6}\Big), \\
{K}_{51}&=U_{51}\sqrt{p_5}\Big(\sqrt{\frac{12}{11}}\ket{1}\bra{1}+\sqrt{\frac{12}{11}}\ket{2}\bra{2}
+\sqrt{\frac{1}{2}}\ket{3}\bra{3}+\ket{4}\bra{4}+\sqrt{\frac{7}{8}}\ket{5}\bra{5}+\sqrt{\frac{10}{7}}\ket{6}\bra{6}\Big),
\end{aligned}\end{eqnarray}
\end{widetext}
we obtain the transformation $\ket{\psi}\bra{\psi}\rightarrow\ket{\eta_1}\bra{\eta_1}$, i.e., $\sum_{i=1}^{5}{K_{i1}}\ket{\psi}\bra{\psi}{K^{\dag}_{i1}}=\ket{\eta_1}\bra{\eta_1}$, where $p_1=1/4$, $p_2=p_3=1/8$, $p_4=1/3$, $p_5=1/6$, $U_{11}=I_6$,  $U_{21}=\ket{1}\leftrightarrow\ket{6}$, $U_{31}=\ket{2}\leftrightarrow\ket{6}$, $U_{41}=\ket{3}\leftrightarrow\ket{5}$, $U_{51}=\ket{3}\leftrightarrow\ket{6}$. We now obtain the transformation $\ket{\eta_1}\bra{\eta_1}\rightarrow\ket{\phi}\bra{\phi}$, where $\ket{\eta_1}$ and $\ket{\phi}$ can be written in the form
\begin{eqnarray}\begin{aligned}
\ket{\eta_1}=\sqrt{\frac{38}{53}}\ket{\chi}
+\sqrt{\frac{15}{53}}\Big(\sqrt{\frac{8}{15}}\ket{4}+\sqrt{\frac{7}{15}}\ket{5}\Big),
\end{aligned}\end{eqnarray}
\begin{eqnarray}\begin{aligned}
\ket{\phi}=\sqrt{\frac{38}{53}}\ket{\chi}
+\sqrt{\frac{15}{53}}\Big(\sqrt{\frac{9}{15}}\ket{4}
+\sqrt{\frac{6}{15}}\ket{5}\Big),
\end{aligned}\end{eqnarray}
respectively. Here, $\ket{\chi}=\sqrt{\frac{6}{19}}\ket{1}+\sqrt{\frac{6}{19}}\ket{2}
+\sqrt{\frac{5}{19}}\ket{3}+\sqrt{\frac{2}{19}}\ket{6}$ is a normalized coherent state. Then, under the Kraus operators given by
\begin{widetext}
\begin{eqnarray}\begin{aligned}
{K}_{12}&=U_{12}\sqrt{p_1}\Big(\ket{1}\bra{1}+\ket{2}\bra{2}+\ket{3}\bra{3}
+\frac{3}{2\sqrt{2}}\ket{4}\bra{4}+\sqrt{\frac{6}{7}}\ket{5}\bra{5}+\ket{6}\bra{6}\Big), \\
{K}_{22}&=U_{12}\sqrt{p_2}\Big(\ket{1}\bra{1}+\ket{2}\bra{2}+\ket{3}\bra{3}
+\frac{\sqrt{3}}{2}\ket{4}\bra{4}+\frac{3}{\sqrt{7}}\ket{5}\bra{5}+\ket{6}\bra{6}\Big),
\end{aligned}\end{eqnarray}
\end{widetext}
we obtain the transformation $\ket{\eta_1}\bra{\eta_1}\rightarrow\ket{\phi}\bra{\phi}$, i.e., $\sum_{i=1}^{2}{K_{i2}}\ket{\eta_1}\bra{\eta_1}{K^{\dag}_{i2}}=\ket{\phi}\bra{\phi}$, where $p_1=2/3$, $p_2=1/3$, $U_{12}=I_6$, $U_{22}=\ket{4}\leftrightarrow\ket{5}$, and  $\sum_{i=1}^{2}{K^{\dag}_{i2}}{K_{i2}}=I_6$. Thus, we obtain the entire transformation $\ket{\psi}\bra{\psi} \rightarrow \ket{\phi}\bra{\phi}$ as $\ket{\psi}\bra{\psi}$ $\rightarrow$ $\ket{\eta_1}\bra{\eta_1}$ $\rightarrow$ $\ket{\phi}\bra{\phi}$, where $\mu{(\psi)}$ $\prec$ $\mu{(\eta_1)}$ $\prec$ $\mu{(\phi)}$. As an example of the nonuniqueness of the intermediate states we may consider another coherent state for which
\begin{eqnarray}\begin{aligned}
\mu{(\tilde{\eta}_1)}= \frac{1}{53}(12, 12, 7, 9, 6, 7)^T.
\end{aligned}\end{eqnarray}
satisfying $\mu{(\psi)}\prec\mu{(\tilde{\eta}_1)}\prec\mu{(\phi)}$.
Hence, it is always possible to find intermediate states satisfying the majorization condition $\mu{(\psi)}$ $\prec$ $\mu{(\eta_1)}$ $\prec$ $\dots$ $\prec$ $\mu{(\eta_k)}$ $\prec$ $\mu{(\phi)}$ and make the transformations $\ket{\psi}\bra{\psi}$ $\rightarrow$ $\ket{\eta_1}\bra{\eta_1}$ $\rightarrow$ $\dots$ $\rightarrow$ $\ket{\eta_k}\bra{\eta_k}$ $\rightarrow$ $\ket{\phi}\bra{\phi}$ using the Kraus operators which can be obtained using protocol I.

\subsection{Deterministic transformations of $d\otimes d$ bipartite entangled pure states}
The results we obtained for coherent state transformations can easily be adapted to $d$-level bipartite entangled pure state transformations (with the majorization criteria satisfied). The initial state $\ket{\Psi}=\sum_{j=1}^{d}\psi_j\ket{j}\ket{j}$ can be transformed to the final state $\ket{\Phi}=\sum_{j=1}^{d}\phi_j\ket{j}\ket{j}$ by performing a $d$-outcome measurement on one of the particles with operators defined by Eq. \eqref{POVM}, where the parameters $c_{sij}$ are the same as found by protocol-I. The POVM measurement yields one of the states
\begin{eqnarray}
M_s^i\ket{\Psi}\rightarrow\ket{\Phi_s^i}=\sum_{j=1}^{d}c_{sij}\ket{j}\ket{j} \quad (i=1,\dots,d), \
\end{eqnarray}
with probabilities $p_s^i=\bra{\Psi}{M_s^i}^{\dag}M_s^i\ket{\Psi}$. Depending on the result of the measurement one of the the unitary transformations $U_s^i$ is applied on both particles to obtain the target state, i.e., $(U_s^i \otimes U_s^i) \ket{\Phi_s^i} =\ket{\Phi}$.

\subsubsection{Examples for entanglement transformations}

We now give three illustrative examples both to demonstrate the adaptability of our results to the deterministic transformations of entanglement and to enhance the intelligibility of our method.

We first consider the complete solution of three-level systems. There are two possible cases for $d=3$: Either $\psi_2 \leq \phi_2$ or $\psi_2 \geq \phi_2$ may exist where $\psi_1 \leq \phi_1$ and $\psi_3 \geq \phi_3$ follow from the majorization condition.

\begin{figure}[!htbp]
		\centering
	\includegraphics[width=0.08\textwidth]{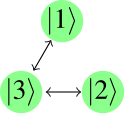}
		\caption{Pictorial representation of the permutations given in Table \ref{table-d3-1}. These two permutations here constitute a
         single SP together with identity transformation $I_3$.}
		\label{figure-n3-1}
\end{figure}

\textit{Example 1: The case $\psi_1 \leq \phi_1$, $\psi_2 \leq \phi_2$, $\psi_3 \geq \phi_3$}. We first construct Table \ref{table-d3-1}.
\begin{table}[!htbp]
\caption{All possible permutations for the case $\psi_1 \leq \phi_1$, $\psi_2 \leq \phi_2$, and $\psi_3 \geq \phi_3$ of $d=3$.}
\label{table-d3-1}
\centering
\begin{ruledtabular}
\begin{tabular}{c c c c}
         & $\psi_1 \leq \phi_1$ & $\psi_2 \leq \phi_2$   \\ [0.5ex] \hline
$\psi_3 \geq \phi_3$ & $\ket{1}\leftrightarrow \ket{3}$ & $\ket{2}\leftrightarrow \ket{3}$
\end{tabular}
\end{ruledtabular}
\end{table}
The permutations in Table \ref{table-d3-1} constitute the SP together with identity transformation $I_3$ (while there are $d-1$ permutations in Table \ref{table-d3-1}, a single SP is sufficient). Figure \ref{figure-n3-1} provides the pictorial representation of permutations given in Table \ref{table-d3-1}.
The SP is then obtained such that
\begin{eqnarray}\label{permut-d3-1-ent}
U=\Big\{U^1, U^2, U^3\Big\}=\Big\{I_3, \ket{2}\leftrightarrow \ket{3}, \ket{1}\leftrightarrow \ket{3}\Big\}.
\end{eqnarray}
The measurement stage begins after completing the first and most important step$-$obtaining the correct set of permutations $U^i$. A generalized three-outcome measurement with the measurement operators,
\begin{eqnarray}\label{povm-d3}
M^i=\sqrt{p^i}\sum_{j=1}^{3}\frac{c_{ij}}{\psi_j}\ket{j}\bra{j}, \quad (\sum_{i=1}^{3} {M^i}^{\dag}M^i=I_3),
\end{eqnarray}
is performed on one of the parties of the initial state $\ket{\Psi}=\sum_{j=1}^{3}\psi_j\ket{j}\ket{j}$. Here, $c_{ij}$ is the $(ij)$th element of the matrix $c$ where $(c_{i1}, c_{i2}, c_{i3})^{T}=(U^i)^{\dag}(\phi_1, \phi_2, \phi_3)^{T}$. The matrix $c$ is given by
\begin{eqnarray}\begin{aligned}
c
=\left(\begin{array}{ccc}  \phi_1 & \phi_2 & \phi_3 \\ \phi_1 & \phi_3 & \phi_2 \\ \phi_3 & \phi_2 & \phi_1 \end{array}\right).
\end{aligned}\end{eqnarray}
The state after the measurement turns out to be one of the states
\begin{eqnarray}\label{yields-d3}
M^i\ket{\Psi}\rightarrow\ket{\Phi^i}=\sum_{j=1}^{3}c_{ij}\ket{j}\ket{j} \quad (i=1,2,3),
\end{eqnarray}
with probabilities $p^i=\bra{\Psi}{M^i}^{\dag}M^i\ket{\Psi}$.
The state $\ket{\Phi^1}$ is already the target state $\ket{\Phi}=\sum_{j=1}^{3}\phi_j\ket{j}\ket{j}$, and the states $\ket{\Phi^2}$ and $\ket{\Phi^3}$ can be transformed to the target state by local unitary transformations (permutations) $\ket{2}\leftrightarrow\ket{3}$ and $\ket{1}\leftrightarrow\ket{3}$, respectively. In other words, the POVM measurement yields one of the states
\begin{eqnarray}\begin{aligned}
\ket{\Phi^1}=&\sum_{j=1}^{3}\phi_j\ket{j}\ket{j}, \\
\ket{\Phi^2}=&\phi_1\ket{1}\ket{1}+\phi_3\ket{2}\ket{2}+\phi_2\ket{3}\ket{3}, \\
\ket{\Phi^3}=&\phi_3\ket{1}\ket{1}+\phi_2\ket{2}\ket{2}+\phi_1\ket{3}\ket{3},
\end{aligned}\end{eqnarray}
with probabilities $p^i$ where $(U^2\otimes U^2)\ket{\Phi^2}=\ket{\Phi}$, $(U^3\otimes U^3)\ket{\Phi^3}=\ket{\Phi}$ and the permutations $U^i$ are given in Eq. \eqref{permut-d3-1-ent}. Also, the condition $\sum_{i=1}^{3} {M^i}^{\dag}M^i=I_3$ implies that the following three linear equations
\begin{eqnarray}\label{xij-d3}
\sum_{i=1}^{3} {p^i} {c^2_{ij}}=\psi^2_j, \quad (j=1,2,3),
\end{eqnarray}
should be satisfied, and solutions of these linear equations give the probabilities.
Thus, probabilities are found to be
\begin{eqnarray}\begin{aligned}
p^1=1-p^2-p^3, \quad
p^2=\frac{\phi_2^2-\psi_2^2}{\phi_2^2-\phi_3^2}
, \quad
p^3=\frac{\phi_1^2-\psi_1^2}{\phi_1^2-\phi_3^2}.
\end{aligned}\end{eqnarray}

\begin{figure}[!htbp]
		\centering
	\includegraphics[width=0.08\textwidth]{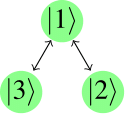}
		\caption{Pictorial representation of the permutations given in Table \ref{table-d3-2}. Here, two permutations constitute a
         single SP together with identity transformation $I_3$.}
		\label{figure-n3-2}
\end{figure}

\textit{Example 2: The case $\psi_1 \leq \phi_1$, $\psi_2 \geq \phi_2$, $\psi_3 \geq \phi_3$}. The permutations in Table \ref{table-d3-2} constitute the SP together with identity transformation $I_3$ (a single SP is sufficient).
\begin{table}[!htbp]
\caption{All possible permutations for the case $\psi_1 \leq \phi_1$, $\psi_2 \geq \phi_2$, and $\psi_3 \geq \phi_3$ of $d=3$.}
\label{table-d3-2}
\centering
\begin{ruledtabular}
\begin{tabular}{c c c c}
         & $\psi_1 \leq \phi_1$    \\ [0.5ex] \hline
$\psi_2 \geq \phi_2$ & $\ket{1}\leftrightarrow \ket{2}$  \\
$\psi_3 \geq \phi_3$ & $\ket{1}\leftrightarrow \ket{3}$
\end{tabular}
\end{ruledtabular}
\end{table}
Figure \ref{figure-n3-2} provides the pictorial representation of permutations given in Table \ref{table-d3-2}.
Then, the SP is obtained such that
\begin{eqnarray}\label{permut-d3-2-ent}
U=\Big\{U^1, U^2, U^3\Big\}=\Big\{I_3, \ket{1}\leftrightarrow \ket{2}, \ket{1}\leftrightarrow \ket{3}\Big\}.
\end{eqnarray}
A generalized three-outcome measurement with the measurement operators defined by Eq. \eqref{povm-d3}
is performed on one of the particles of the initial state $\ket{\Psi}$. In Eq. \eqref{povm-d3}, $c_{ij}$ is the $(ij)$th element of the matrix $c$ where $(c_{i1}, c_{i2}, c_{i3})^{T}=(U^i)^{\dag}(\phi_1, \phi_2, \phi_3)^{T}$. The matrix $c$ is given by
\begin{eqnarray}\begin{aligned}
c
=\left(\begin{array}{ccc}  \phi_1 & \phi_2 & \phi_3 \\ \phi_2 & \phi_1 & \phi_3 \\ \phi_3 & \phi_2 & \phi_1 \end{array}\right).
\end{aligned}\end{eqnarray}The state after the measurement turns out to be one of the states, given in Eq. \eqref{yields-d3}, with probabilities $p^i=\bra{\Psi}{M^i}^{\dag}M^i\ket{\Psi}$.
The state $\ket{\Phi^1}$ is already the target state $\ket{\Phi}$, and the states $\ket{\Phi^2}$ and $\ket{\Phi^3}$ can be transformed to the target state by local unitary transformations (permutations) $\ket{1}\leftrightarrow\ket{2}$ and $\ket{1}\leftrightarrow\ket{3}$, respectively. In other words, the POVM measurement yields one of the states
\begin{eqnarray}\begin{aligned}
\ket{\Phi^1}=&\sum_{j=1}^{3}\phi_j\ket{j}\ket{j}, \\
\ket{\Phi^2}=&\phi_2\ket{1}\ket{1}+\phi_1\ket{2}\ket{2}+\phi_3\ket{3}\ket{3}, \\
\ket{\Phi^3}=&\phi_3\ket{1}\ket{1}+\phi_2\ket{2}\ket{2}+\phi_1\ket{3}\ket{3},
\end{aligned}\end{eqnarray}
with probabilities $p^i$ where $(U^2\otimes U^2)\ket{\Phi^2}=\ket{\Phi}$, $(U^3\otimes U^3)\ket{\Phi^3}=\ket{\Phi}$ and the permutations $U^i$ are given in Eq. \eqref{permut-d3-2-ent}.
The condition $\sum_{i=1}^{3} {M^i}^{\dag}M^i=I_3$ implies that the following linear equations
\begin{eqnarray}\label{xij-d3}
\sum_{i=1}^{3} {p^i} {c^2_{ij}}=\psi^2_j, \quad (j=1,2,3),
\end{eqnarray}
should be satisfied, and solutions of these linear equations give the probabilities.
Thus, probabilities are found to be
\begin{eqnarray}\begin{aligned}
p^1=1-p^2-p^3, \quad
p^2=\frac{\psi_2^2-\phi_2^2}{\phi_1^2-\phi_2^2}
, \quad
p^3=\frac{\psi_3^2-\phi_3^2}{\phi_1^2-\phi_3^2}.
\end{aligned}\end{eqnarray}
We eventually obtained the complete solution for $d=3$ by solving the two possible cases.

\begin{figure}[!htbp]
		\centering
	\includegraphics[width=0.09\textwidth]{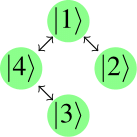}
		\caption{Pictorial representation of the permutations given in Table \ref{table-d4-ent}. Here, three permutations constitute a single SP together with identity transformation $I_4$.}
		\label{figure-n4-ent}
\end{figure}

\textit{Example 3.} As a third example, we consider the case $\psi_2 \geq \phi_2$ and $\psi_3 \leq \phi_3$ for $d=4$ where $\psi_1 \leq \phi_1$ and $\psi_4 \geq \phi_4$ follow from the majorization condition.  To begin with, we construct Table \ref{table-d4-ent}.
\begin{table}[!htbp]
\caption{All possible permutations for the case $\psi_1 \leq \phi_1$, $\psi_2 \geq \phi_2$, $\psi_3 \leq \phi_3$, and $\psi_4 \geq \phi_4$  of $d=4$.}
\label{table-d4-ent}
\centering
\begin{ruledtabular}
\begin{tabular}{c c c c}
         & $\psi_1 \leq \phi_1$ & $\psi_3 \leq \phi_3$   \\ [0.5ex] \hline
$\psi_2 \geq \phi_2$ & $\ket{1}\leftrightarrow \ket{2}$ & - \\
$\psi_4 \geq \phi_4$ & $\ket{1}\leftrightarrow \ket{4}$ & $\ket{3}\leftrightarrow \ket{4}$
\end{tabular}
\end{ruledtabular}
\end{table}
The permutations in Table \ref{table-d4-ent} constitute the SP together with identity transformation $I_4$ (while there are $d-1$ permutations in Table \ref{table-d4-ent}, a single SP is sufficient). Figure \ref{figure-n4-ent} provides the pictorial representation of permutations given in Table \ref{table-d4-ent}.
The SP is then obtained such that
\begin{eqnarray}\begin{aligned}\label{permut-d4-ent}
U=&\Big\{U^1, U^2, U^3, U^4\Big\} \\
=&\Big\{I_4, \ket{1}\leftrightarrow \ket{2}, \ket{3}\leftrightarrow \ket{4},
\ket{1}\leftrightarrow \ket{4}\Big\}.
\end{aligned}\end{eqnarray}
A generalized four-outcome measurement with the measurement operators,
\begin{eqnarray}\label{povm-d4}
M^i=\sqrt{p^i}\sum_{j=1}^{4}\frac{c_{ij}}{\psi_j}\ket{j}\bra{j}, \quad (\sum_{i=1}^{4} {M^i}^{\dag}M^i=I_4),
\end{eqnarray}
is performed on one of the particles of the initial state $\ket{\Psi}=\sum_{j=1}^{4}\psi_j\ket{j}\ket{j}$. Here, $c_{ij}$ is the $(ij)$th element of the matrix $c$ where $(c_{i1}, c_{i2}, c_{i3}, c_{i4})^{T}=(U^i)^{\dag}(\phi_1, \phi_2, \phi_3, \phi_4)^{T}$. The matrix $c$ is given by
\begin{eqnarray}\begin{aligned}
c
=\left(\begin{array}{cccc}  \phi_1 & \phi_2 & \phi_3 & \phi_4\\ \phi_2 & \phi_1 & \phi_3 & \phi_4 \\ \phi_1 & \phi_2 & \phi_4 & \phi_3 \\ \phi_4 & \phi_2 & \phi_3 & \phi_1 \end{array}\right).
\end{aligned}\end{eqnarray}
The state after the measurement turns out to be one of the states
\begin{equation}\label{yields-d4}
M^i\ket{\Psi}\rightarrow\ket{\Phi^i}=\sum_{j=1}^{4}c_{ij}\ket{j}\ket{j} \quad (i=1,2,3,4),
\end{equation}
with probabilities $p^i=\bra{\Psi}{M^i}^{\dag}M^i\ket{\Psi}$.
The state $\ket{\Phi^1}$ is already the target state $\ket{\Phi}=\sum_{j=1}^{4}\phi_j\ket{j}\ket{j}$, and the states $\ket{\Phi^2}$, $\ket{\Phi^3}$ and $\ket{\Phi^4}$ can be transformed to the target state by local unitary transformations (permutations) $\ket{1}\leftrightarrow\ket{2}$, $\ket{3}\leftrightarrow\ket{4}$ and $\ket{1}\leftrightarrow\ket{4}$, respectively. Also, the condition $\sum_{i=1}^{4} {M^i}^{\dag}M^i=I_4$ implies that the following linear equations
\begin{eqnarray}\label{xij-d4-ent}
\sum_{i=1}^{4} {p^i} {c^2_{ij}}=\psi^2_j, \quad (j=1,\dots,4),
\end{eqnarray}
should be satisfied, and solutions of these linear equations give the probabilities.
Thus, probabilities are found to be
\begin{eqnarray}\begin{aligned}\label{proba-d4-ent}
p^1&=1-\sum_{i=2}^{4} p^i, \quad
p^2=\frac{\psi_2^2-\phi_2^2}{\phi_1^2-\phi_2^2}
, \quad
p^3=\frac{\phi_3^2-\psi_3^2}{\phi_3^2-\phi_4^2}
, \\
p^4&=\frac{\psi_3^2+\psi_4^2-\phi_3^2-\phi_4^2}{\phi_1^2-\phi_4^2}.
\end{aligned}\end{eqnarray}
Succinctly, the POVM measurement yields one of the states
\begin{eqnarray}\begin{aligned}
\ket{\Phi^1}=&\sum_{j=1}^{4}\phi_j\ket{j}\ket{j}, \\
\ket{\Phi^2}=&\phi_2\ket{1}\ket{1}+\phi_1\ket{2}\ket{2}+\phi_3\ket{3}\ket{3}+\phi_4\ket{4}\ket{4}, \\
\ket{\Phi^3}=&\phi_1\ket{1}\ket{1}+\phi_2\ket{2}\ket{2}+\phi_4\ket{3}\ket{3}+\phi_3\ket{4}\ket{4}, \\
\ket{\Phi^4}=&\phi_4\ket{1}\ket{1}+\phi_2\ket{2}\ket{2}+\phi_3\ket{3}\ket{3}+\phi_1\ket{4}\ket{4},
\end{aligned}\end{eqnarray}
with the probabilities $p^i$ given in Eq. \eqref{proba-d4-ent} where $(U^i\otimes U^i)\ket{\Phi^i}=\ket{\Phi}$ ($i=1,2,3,4$) and the permutations $U^i$ are given in Eq. \eqref{permut-d4-ent}. Since all states obtained after the measurement are transformed to the target state by local unitary transformations and the total probability of success is unity ($\sum_{i=1}^{4} p^i=1$), we may conclude that the deterministic transformation $\ket{\Psi}\rightarrow\ket{\Phi}$ can be obtained by LOCC.


\section{Conclusion}

In this paper, we present two explicit protocols for the deterministic transformations of coherent states under incoherent operations. We first present a permutation-based protocol (protocol I) which reduces the problem of $d$-level deterministic coherence transformations to solving $d$ linear equations for $d$ unknowns (probabilities). We give two illustrative examples to make the protocol clearer. One of the most significant points of protocol I is that it provides a single map, that is, $\Phi_s[\rho_{\psi}]=\rho_{\phi}$. Using protocol I, one can easily find the sets of permutations, probabilities and Kraus operators for $d$-level systems. We then present generalized solutions for some source and target states using protocol I.
We also present an alternative protocol (protocol II) where we use lower dimensional ($d'<d$) solutions of the permutation-based protocol to obtain the complete transformation as a sequence of coherent-state transformations. In each step of the complete transformation, we obtain at least $d'-1$ coefficients of the final coherent state $\ket{\phi}$ using $d'$-dimensional subspace solutions. Thus, we obtain the $d$-level pure coherent state transformations by cascading a sequence
of coherence transformations with the set of incoherent operations $\{\Phi_{(i)}\}_{i=1,2,\dots,\lfloor {(d+d'-3)}/{(d'-1)} \rfloor}$. We discuss an example for protocol II. Both protocols also provide complete solutions for LOCC deterministic transformations of $d \otimes d$ bipartite entangled pure states.


\section{Acknowledgments}
G. T. acknowledges from the Scientific and Technological Research Council of Turkey (TUBITAK) for financial support. We thank G. Adesso and L. Lami for useful discussions and comments.



\begin{thebibliography} {99}
	
\bibitem{qthermody1} F. G. S. L. Brand${\tilde{\text{a}}}$o, M. Horodecki, J. Oppenheim, J. M. Renes, and R.W. Spekkens, Resource Theory
                     of Quantum States Out of Thermal Equilibrium,
                     Phys. Rev. Lett. {\bf111}, 250404 (2013).	


\bibitem{qthermody2} G. Gour, M. P. Müller, V. Narasimhachar, R.W. Spekkens, and N. Y. Halpern, The resource theory of informational
                     nonequilibrium in thermodynamics,
                     Phys. Rep. {\bf583}, 1-58 (2015).


\bibitem{qthermody3} J. Goold, M. Huber, A. Riera, L. del Rio, and P. Skrzypczyk, The role of quantum information in
                     thermodynamics-a topical review,
                     J. Phys. A: Math. Theor. {\bf49}, 143001 (2016).


\bibitem{qthermody4} M. Lostaglio, K. Korzekwa, D. Jennings, and T. Rudolph, Quantum Coherence, Time-Translation Symmetry, and
                     Thermodynamics,
                     Phys. Rev. X {\bf5}, 021001 (2015).	


\bibitem{qmetrology1} V. Giovannetti, S. Lloyd, and L. Maccone, Quantum Metrology,
                      Phys. Rev. Lett. {\bf96}, 010401 (2006).


\bibitem{qmetrology2} G. Tóth and I. Apellaniz, Quantum metrology from a quantum information science perspective,
                      J. Phys. A: Math. Theor. {\bf47}, 424006 (2014)


\bibitem{qmetrology3} N. Friis, D. Orsucci, M. Skotiniotis, P. Sekatski, V. Dunjko, H. J. Briegel, and W. Dür, Flexible resources for
                      quantum metrology,
                      New J. Phys. {\bf19}, 063044 (2017).


\bibitem{qmetrology4} A. R. Shlyakhov, V. V. Zemlyanov, M. V. Suslov, A. V. Lebedev, G. S. Paraoanu, G. B. Lesovik, and G. Blatter,
                      Quantum metrology with a transmon qutrit,
                      Phys. Rev. A {\bf97}, 022115 (2018).


\bibitem{qalgorithms} M. Hillery, Coherence as a resource in decision problems: The Deutsch-Jozsa algorithm and a variation,
                      Phys. Rev. A {\bf93}, 012111 (2016).


\bibitem{qalgorithms1} H. L. Shi, S. Y. Liu, X. H. Wang, W. L. Yang, Z. Y. Yang, and H. Fan, Coherence depletion in the Grover quantum
                       search algorithm,
                       Phys. Rev. A {\bf95},  032307 (2017).


\bibitem{qalgorithms2} S. Chin, Coherence number as a discrete quantum resource,
                       Phys. Rev. A {\bf96},  042336 (2017).




\bibitem{qcoherence} T. Baumgratz, M. Cramer, and M. B. Plenio, Quantifying Coherence,
                     Phys. Rev. Lett. {\bf113}, 140401 (2014).


\bibitem{resourcethercoh} A. Streltsov, G. Adesso, and M. B. Plenio, Colloquium: Quantum coherence as a resource,
                          Rev. Mod. Phys. {\bf89}, 041003 (2017).


\bibitem{Streltsovrtc} A. Streltsov, S. Rana, P. Boes, and J. Eisert, Structure of the Resource Theory of Quantum Coherence,
                       Phys. Rev. Lett. {\bf119}, 140402 (2017).


\bibitem{Dana} K. B. Dana, M. G. D\'{\i}az, M. Mejatty, and A. Winter, Resource theory of coherence: Beyond states,
               Phys. Rev. A {\bf95}, 062327 (2017).


\bibitem{genuine} J. l de Vicente and A. Streltsov, Genuine quantum coherence,
                  J. Phys. A: Math. Theor. {\bf50}, 045301 (2017).


\bibitem{Yang} S. Yang and C. Yu, Operational resource theory of total quantum coherence,
               Annals of Phys. {\bf388}, 305-314 (2018).


\bibitem{Winter} A. Winter and D. Yang, Operational Resource Theory of Coherence,
                 Phys. Rev. Lett. {\bf116}, 120404 (2016).


\bibitem{Alexrtofc} A. Streltsov, S. Rana, M. N. Bera, and M. Lewenstein, Towards Resource Theory of Coherence in Distributed Scenarios,
                    Phys, Rev. X {\bf7}, 011024 (2017).


\bibitem{entanglement} R. Horodecki, P. Horodecki, M. Horodecki, and K. Horodecki, Quantum entanglement,
                       Rev. Mod. Phys. {\bf81}, 865 (2009).


\bibitem{condition} M. A. Nielsen, Conditions for a Class of Entanglement Transformations,
                    Phys. Rev. Lett. {\bf83}, 436 (1999).


\bibitem{vidal} G. Vidal, Entanglement of Pure States for a Single Copy,
                Phys. Rev. Lett. {\bf83}, 1046 (1999).


\bibitem{NielsenVidal} M. Nielsen and G. Vidal, Majorization and the interconversion of bipartite states,
                       Quantum Info. Comput. {\bf1}, 76-93 (2001).


\bibitem{ccoherence} S. Du, Z. Bai and Y. Guo, Conditions for coherence transformations under incoherent operations,
                     Phys. Rev. A {\bf91}, 052120 (2015).


\bibitem{ccoherenceerr} S. Du, Z. Bai and Y. Guo, Erratum: Conditions for coherence transformations under incoherent operations,
                        Phys. Rev. A {\bf95}, 029901(E) (2017).


\bibitem{catalyticcoherence} J. \AA{}berg, Catalytic Coherence,
                             Phys, Rev. Lett. {\bf113}, 150402 (2014).


\bibitem{ccoherence2} S. Du, Z. Bai, and X. Qi, Coherence Measures and Optimal Conversion for Coherent States,
                      Quantum Info. Comput. {\bf15-16}, 1307-1316 (2015).


\bibitem{Chitambar1} E. Chitambar and G. Gour, Comparison of incoherent operations and measures of coherence,
                     Phys, Rev. A {\bf94}, 052336 (2016).


\bibitem{Chitambar} E. Chitambar and G. Gour, Critical Examination of Incoherent Operations and a Physically Consistent Resource
                    Theory of Quantum Coherence,
                    Phys, Rev. Lett. {\bf117}, 030401 (2016).


\bibitem{coherence-vector} H. Zhu, Z. Ma, Z. Cao, S.M. Fei, and V. Vedral, Operational one-to-one mapping between coherence and
                           entanglement measures,
                           Phys. Rev. A {\bf96}, 032316 (2017).


\bibitem{Bhatia} R. Bhatia, \textit{Matrix Analysis} (Springer-Verlag, New York, 1997).


\bibitem{deterministic} G. Torun and A. Yildiz, Deterministic transformations of bipartite pure states,
                        Phys. Lett. A {\bf379}, 113 (2015).

\end{thebibliography}
\end{document}